# Threats and Corrective Measures for IoT Security with Observance of Cybercrime: A Survey

***Sita Rani[a], Aman Kataria[b], Vishal Sharma[c], Smarajit Ghosh[d], Vinod Karar[e], Kyungroul Lee[f], and Chang Choi[g]***

*[a]Department of Computer Science & Engineering, Gulzar Group of Institutes, Khanna-141401, Punjab, INDIA.*
*[b]Optical Devices and Systems, CSIR, CSIO, Chandigarh-160030, INDIA*
*[c*] EEECS, Queen's University Belfast, Northern Ireland, UK*
*[d] Department of Electrical and Instrumentation Engineering, Thapar Institute of Engineering and Technology, Patiala, INDIA.*
*[e]Chief Scientist, Optical Devices and Systems, CSIR-CSIO, Chandigarh-160030, INDIA.*
*[f]School of Computer Software, Daegu Catholic University, Gyeongsan-si, KOREA*
*[g]Department of Computer Engineering, Gachon University, Seongnam, KOREA*

---

**A B S T R A C T**

Internet of Things (IoT) is the utmost assuring framework to facilitate human life with quality and comfort. IoT has contributed significantly to numerous application areas. The stormy expansion of smart devices and their credence for data transfer using wireless mechanics boosts their susceptibility to cyber-attacks. Consequently, the rate of cybercrime is increasing day by day. Hence, the study of IoT security threats and possible corrective measures can benefit the researchers to identify appropriate solutions to deal with various challenges in cybercrime investigation. IoT forensics plays a vital role in cybercrime investigations. This review paper presents an overview of the IoT framework consisting of IoT architecture, protocols, and technologies. Various security issues at each layer and corrective measures are also discussed in detail. This paper also presents the role of IoT forensics in cybercrime investigation in various domains like smart homes, smart cities, automated vehicles, healthcare, etc. Along with, the role of advanced technologies like Artificial Intelligence, Machine Learning, Cloud computing, Edge computing, Fog computing, and Blockchain technology in cybercrime investigation are also discussed. At last, various open research challenges in IoT to assist cybercrime investigation are explained to provide a new direction for further research.

## 1. Introduction

The term 'Internet of Things' (IoT) characterizes the network of devices – "things"- which are equipped with different types of sensors, advanced technologies, and software. Although the concept of IoT was introduced by Kevin Aston in the year 1999, but in the last few years it developed very briskly and became one of the most prominent technologies of the era [1, 2]. Smart devices and things have the features to gather process and communicate data to deliver several services and applications for the convenience of the users. [3], [4], and [5]. Consequently, it is not a single technology but a strong merger of 5G and Beyond, Big Data, Artificial Intelligence, Edge Computing, FinTech, and Cloud Computing [6], as shown in Figure 1 which represents IoT as the conflux of technologies.

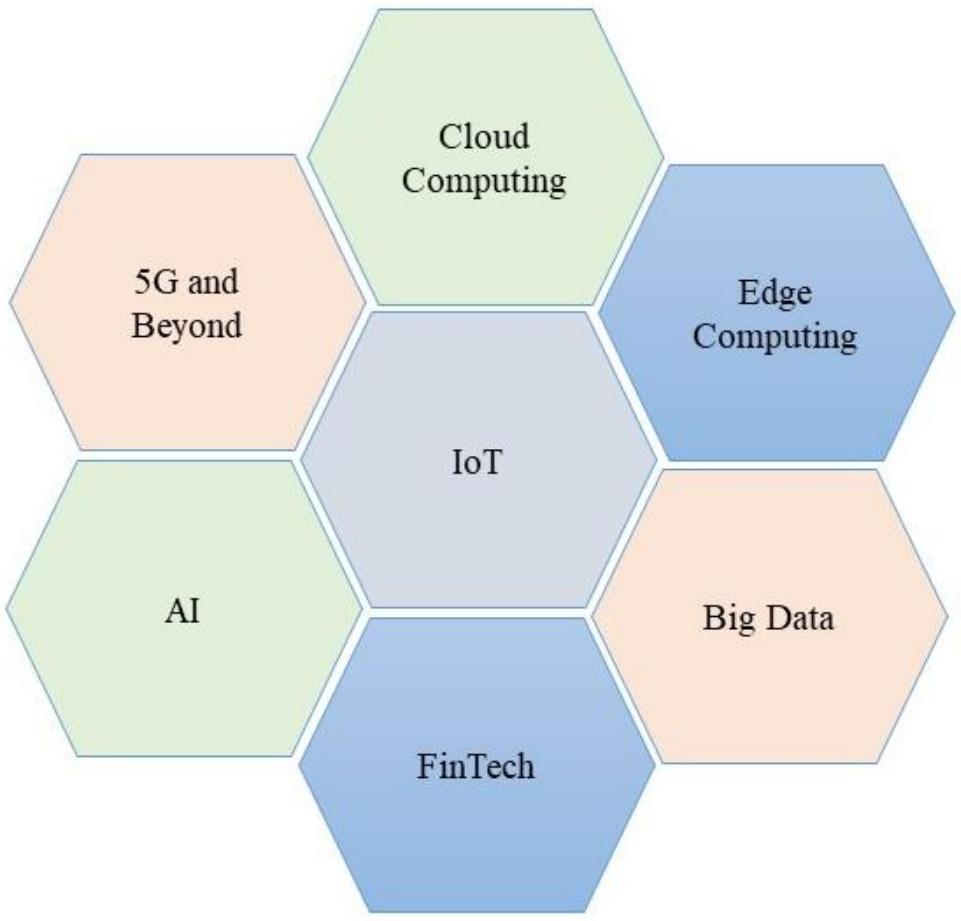

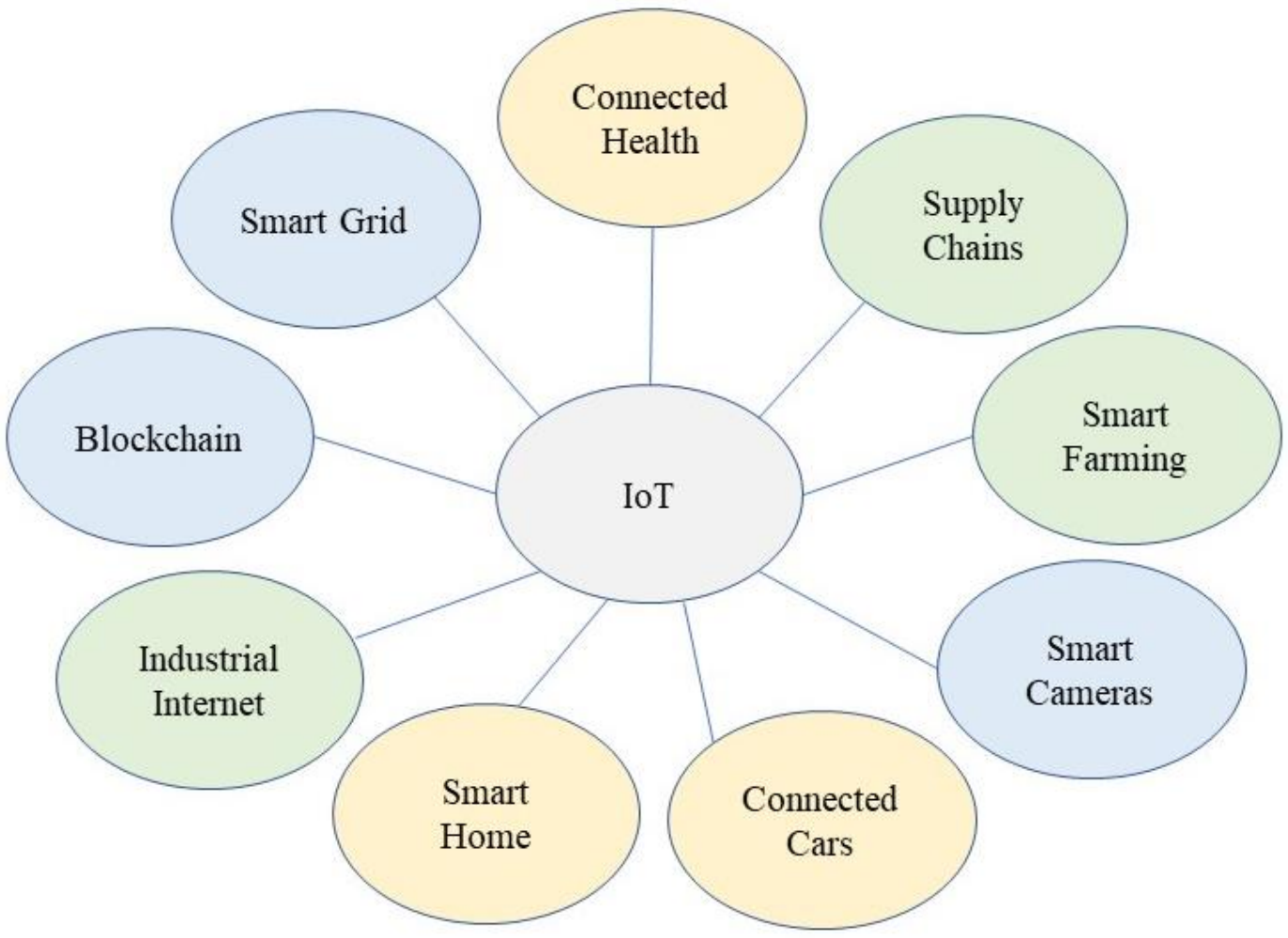

**Figure 1- IoT: A conflux of technologies [1, 3, 5, 6]**

**Figure 2- Different applications of IoT [1, 6-12]**

In a short period, IoT has been deployed in many domains. Their applications range from simple household devices to very complex and sophisticated industrial equipment and machines. Smart healthcare, supply chains, smart farming, unmanned vehicles, smart homes, underwater IoT sensors, smart cars, smart grids, and smart industry are some of the most benefitted areas through IoT, shown in Figure 2 [13],[14]. IoT has also confiscated a wide range of objects to provide more lifestyle-friendly digitized services [15].

As all the smart devices are connected through cyberspace, so increase in their number also widens the surface for cybercrime. Although the domain of cybersecurity benefitted from the involvement of IoT devices but in-parallel introduced different types of security issues[16]. A sharp hike is observed in the statistics of security attacks and cybercrimes across the world from the reports published by the Internet Crime Complaint Centre (IC3) in the year 2019, shown in Table 1. From the year 2015 to 2020, a total of 3,919,014 complaints have been received, which caused a total loss of $23.5 Billion. As per the facts published by IC3, India is 3$^{rd}$ in the list of top 20 crime victim countries [17].

**Table 1- Data on crime complaints and financial losses from 2015 to 2020 [18, 19]**

| Year | Number of complaints Registered | Total Loss (In $Billion) |
|---|---|---|
| **2015** | 288012 | 1.1 |
| **2016** | 298728 | 1.5 |
| **2017** | 301580 | 1.4 |
| **2018** | 351937 | 2.7 |
| **2019** | 467361 | 3.5 |
| **2020** | 2211396 | 13.3 |

The era of IoT-enabled devices is blooming expeditiously. This rapid development is introducing both opportunities and obstacles for the identification of physical and cyber threats [20]. These attacks are malignant actions planned to damage significant data and information and to disturb important services [21], [22] in different types of IoT devices equipped with sensors. [23]. Although IoT – enabled devices facilitate the process of cybercrime detection but are themselves prone to cyber threats. One workable security solution lies at the manufacturer's end. At the time of design and development of smart devices and applications, secure technologies and protocols are required to be practiced. But IoT-enabled devices provide an increased surface for cyber threats due to indigent security measures. Security threats are creating a large torment for the versatile IoT systems. The level of the security threats in the IoT domain may be even life threatening.

Data from the main academic databases have been collected to study the scope of potential research in the domain of cybersecurity in IoT [18]. Figure 3 depicts the number of research papers referred to in the survey related to security issues in IoT from the year 1998 to 2020. As analyzed, this area of research has gained a lot of importance in the last decade.

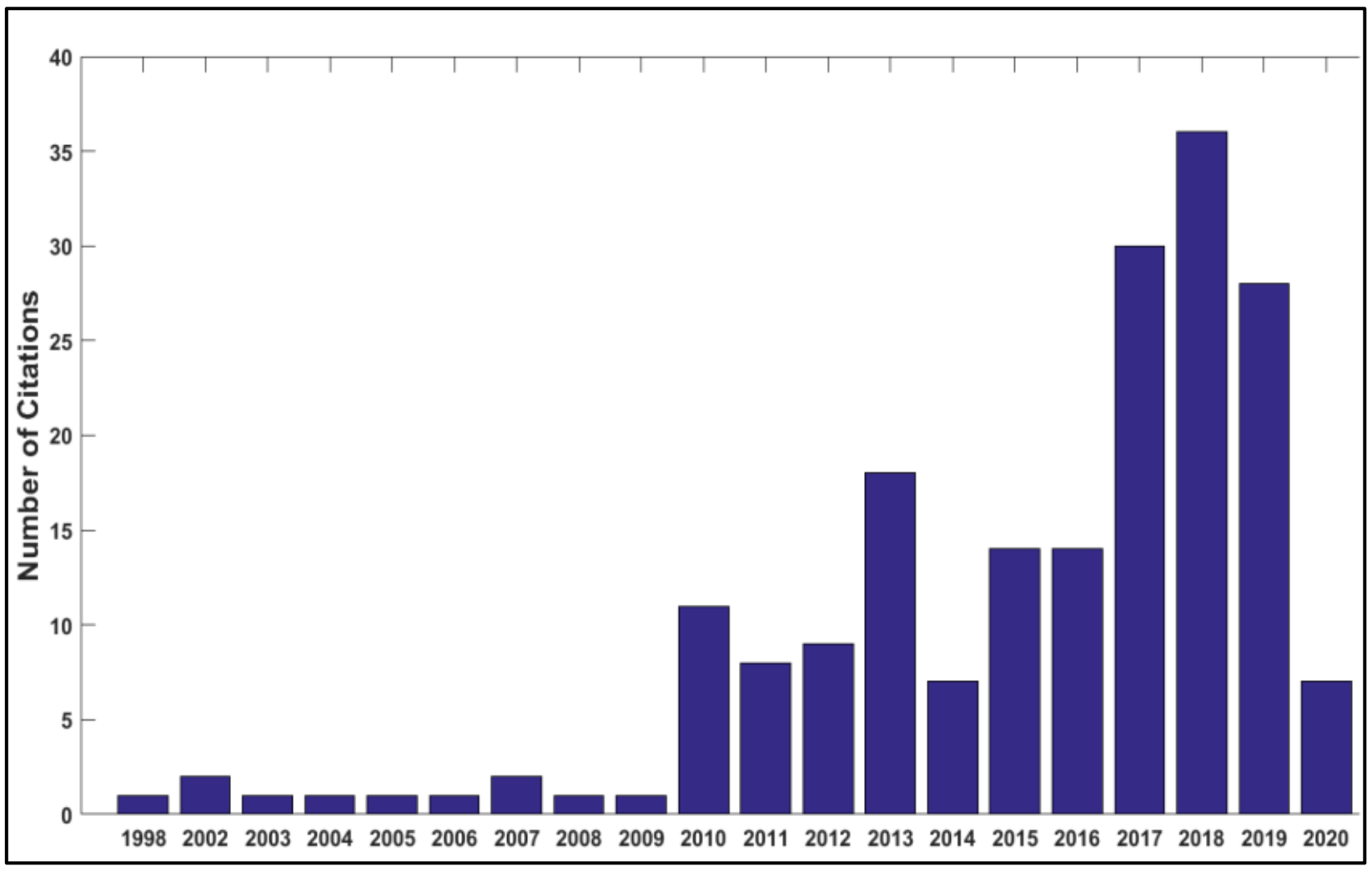

**Figure 3 - Number of articles cited in this survey from the year 1998 to 2020**

**Table 2- Comparison of features of this survey with the existing survey articles**

| S.No. | Author | Year | Cyber-attacks and security | Cyber Crime | Security in IoT devices | Privacy in IoT devices | Patents reported | Discussions on real-time applications |
|---|---|---|---|---|---|---|---|---|
| 1 | Burhan *et al.* | 2018 | ✘ | ✘ | ✓ | ✓ | ✘ | ✘ |
| 2 | Williams *et al.* | 2018 | ✓ | ✓ | ✘ | ✘ | ✘ | ✘ |
| 3 | Huang *et al.* | 2018 | ✓ | ✓ | ✘ | ✘ | ✘ | ✘ |
| 4 | Bhat and Dutta | 2019 | ✓ | ✘ | ✓ | ✘ | ✘ | ✓ |
| 5 | Mrabet *et al.* | 2018 | ✓ | ✘ | ✘ | ✘ | ✘ | ✘ |
| 6 | Tounsi and Rais | 2017 | ✓ | ✓ | ✘ | ✘ | ✘ | ✓ |
| 7 | Jian-hua Li | 2018 | ✓ | ✘ | ✘ | ✘ | ✘ | ✘ |
| 8 | Khadam *et al.* | 2020 | ✓ | ✘ | ✓ | ✓ | ✘ | ✘ |
| 9 | Shafiq *et al.* | 2018 | ✓ | ✘ | ✘ | ✘ | ✘ | ✘ |
| 10 | Weichbroth and Lysik | 2020 | ✓ | ✓ | ✓ | ✘ | ✘ | ✘ |
| 11 | This survey | 2021 | ✓ | ✓ | ✓ | ✓ | ✓ | ✓ |

Keeping in view the relevance of the domain and need of the hour, in this paper, we discuss IoT architecture, security systems, and potential IoT security threats that may cause cybercrime to occur. Along with, IoT forensics and its contribution to crime investigation are also discussed in detail. Table 2 presents the merits of this survey in comparison to the existing latest surveys. It visualizes the novelty of this survey as much emphasis is focussed on cybercrime, Patents reported and real-time applications developed to mitigate the problems occurring due to cybercrime in IoT devices.

In the remainder of this paper, Section 2 is focused on the various types of risks associated with the IoT environment. Existing work on IoT security and cybercrime and the scope of this survey are presented in Section 3. Section 4 is focused on IoT framework and applications. The role of digital forensics in cybercrime investigation is elaborated in Section 5. Section 6 presents the role of advanced technologies in IoT security. Open research challenges in the domain to provide a new direction to the researchers are discussed in Section 7 and the paper is concluded in Section 8.

## 2. Risks in IoT

The IoT evolution is prone to cause a diversity of ethical problems in society like unauthorized access to confidential information, privacy breach, misuse of secret data, identity theft, etc. Although these problems were already existing in the era of the internet and Information and Communication Technology (ICT) but became more dominant in IoT systems [24]. Figure 4 describes several potential risks associated with IoT.

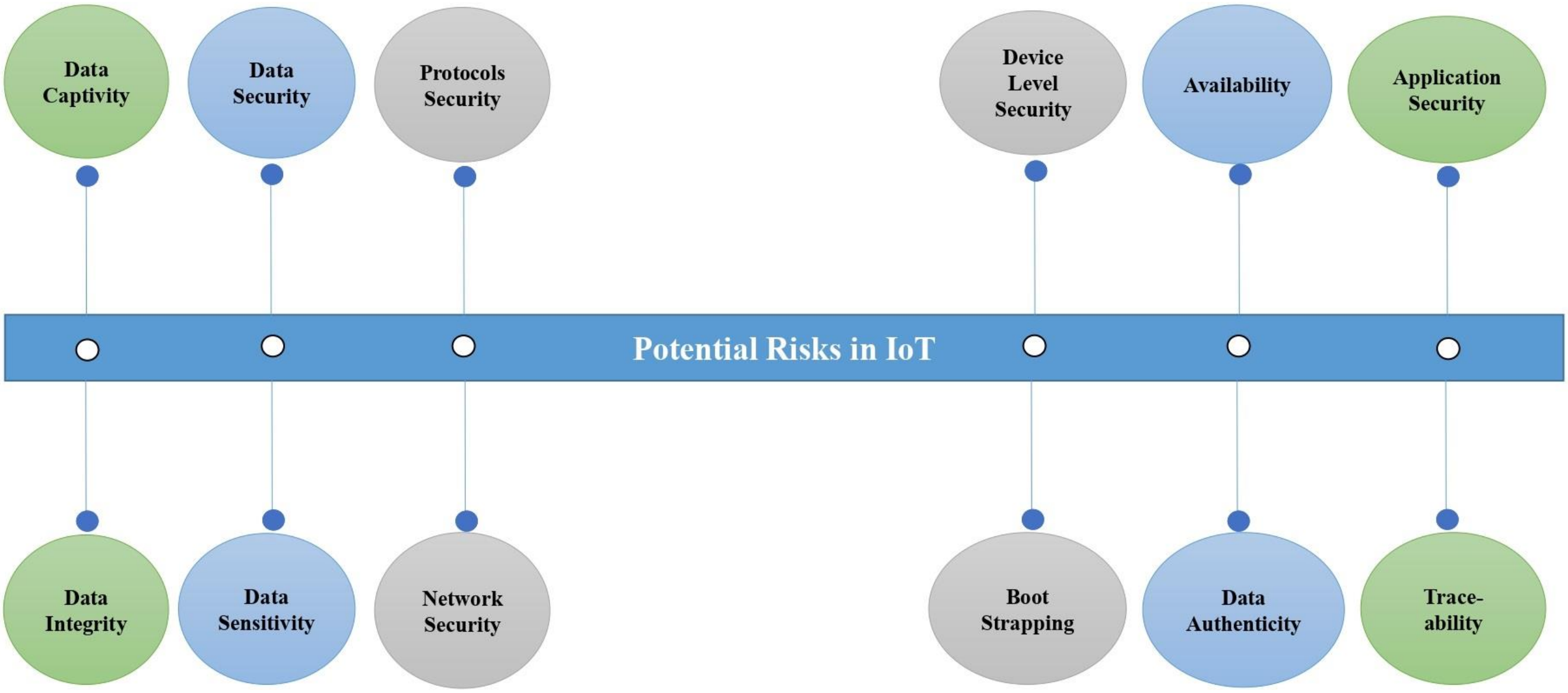

**Figure 4 - Risks of IoT [24-29]**

### *2.1 Privacy Facet*

Confidentiality of users and secrecy of the data generated from numerous business processes are the major areas of concern linked to the IoT [30]. The dominant usage of versatile devices with poor security mechanisms leads to mismanagement of the IoT system [31]. To handle the security issues related to data generated by the IoT devices, there is a requirement for advanced cryptography techniques. However, these techniques should be energy-optimized and have the potential to synchronize with the dynamism of smart devices [32]. With the advancement in IoT, many privacy issues evolved [24], [33].

- **Data Captivity:** A few moralistic questions related to the user data remain unanswered, such as generating unlawful leverage, hard competition, etc. These issues are essential to evade consumer captivity through data [36].
- **Data Integrity:** The consistency and accuracy of the data are the primary requirements for the integrity of the data in IoT devices. Maintaining data integrity is the main motive of enterprise security solutions as compromising data integrity can lead to the loss of sensitive data. Data integrity is necessary for reusability, searchability, recovery, and traceability of the data.
- **Data security:** The data must be secured from illegitimate usage on the devices as well as during transmission in the IoT environment. The diversity of the IoT devices and different communication modes cause a challenge for data security protocols, which is the root cause of security breaches [35]. Another major threat to data security is the variety of applications using this data and cause personal information of the user exposed to cyber-attacks.
- **Data Sensitivity:** Several applications collect the user's personal information sometimes even without knowledge. Therefore, the sensitivity of the data is a major area of concern. The major risks associated with this data are the frame of reference of usage of this data. Consequently, there should be some security protocols for context-aware data collection and usage [34].
- **Protocols Security:** Because of the versatility of devices and collaborators convoluted in the stationing of IoT, the biggest challenge is the applicability of law and regulations for the formation of authentic protocols for communication in the IoT. As the IoT systems are evolving day by day and becoming global; so, there is a clear possibility of the applicability of multiple legislations. Besides, this is an important area for awareness among users, IoT manufacturers, and law builders.
- **Network Security:** The network plays a vital role in the security of IoT devices. The IoT device is connected with the network for the data and workload. This data can become an easy target for hackers or attackers that can compromise the whole system. It is necessary to adapt and devise effective methods to protect the network with which the IoT device is connected.
- **Device Level Security:** The security of an IoT device is considered at the beginning of its design. To ensure the secure implementation of an IoT device, a secure architecture is deployed. During the manufacturing of an IoT device, the care is taken in terms of secure digital device IDs. The credentials are trusted to tackle against various attacks like data and device cloning, data tempering or any other misuse.
- **Boot Strapping:** Bootstrapping refers to any process which occurs before any IoT device becomes operational. Bootstrapping is necessary for the IoT devices of the present generation. The time of bootstrapping in the initial configuration of an IoT device plays a vital role. Therefore the bootstrapping process in IoT devices should be highly secure process.
- **Availability**: The blend of IoT in services related to health, security, etc. has caused the continuous availability of these services a critical issue. Many people are heavily dependent on the IoT devices utilized to provide these services. Therefore, any loss to these services will severely impact human life.
- **Data Authenticity:** Ownership of the collected user data is a major unaddressed issue in IoT along with data management. Once the user stops using the service, the personal information remains with the service provider and can be sold further to generate revenue.
- **Application Security:** The applications designed for various IoT devices are also vulnerable to different types of attacks. It is necessary that the application should be secure and defensive in nature to counter attack the attackers and malwares. There are different types of attacks which can intrude in the architecture of an IoT device like DDoS, spam attack, Message interception through a spyware, vulnerable 3pp library and injection attacks.
- **Traceability.** In an IoT environment, users must have the right to pass consent to provide personal information to numerous real-life services. The implemented security protocols and mechanisms should ensure user identification on the network, but restrict the user traceability to attackers from personal information [37], [38].

### *2.2 Security Facet*

The security of a computer system encompasses various methods and techniques that safeguard all kinds of resources from illegitimate access. Resources may include hardware, software and data, whereas illegitimate access may include unauthorized usage or damage to resources. In IoT systems, security aspects focus on architecture, the security model of every device, bootstrapping, network security and application security [39]. Security architecture demonstrates the various system components liable to ensure the security of an IoT device. The security model of each device focuses on the implementation of security methods and criteria along with the management of various applications. Network security deals with the reliable functioning of IoT. Online application security is all about the authentication of various things on the network for communication and exchange of data. Network security is highly dependent on the internet, which is an anxious media of data exchange and leads to a large possibility of data-stealing. The deployment of IoT is dependent on the internet and computer networks. Consequently, it is affected by all security issues related to computer networks as well as the internet. Before using IoT devices, all stakeholders should analyze the associated risks related to the security and privacy of the user information. Accordingly, more sophisticated security policies must be designed by governing organizations.

### *2.3 Cybercrime*

Like any other crime, cybercrime may have a variety of aspects and may be committed in different plots. Several definitions of cybercrime are available, given from different aspects i.e. sufferers, protector, or viewer. According to the definition given by Newman [40], cybercrime is an action in which computers or computer networks are used as a means, purpose, or platform to execute some criminal act. It may consist of some information theft or usage of computers to do some other criminal activity. The Council of Europe's Cybercrime Treaty defines cybercrime as any act of data content or copyright transgression. The 'Manual on the Prevention and Control of Computer-Related Crime' by the United Nations defines cybercrime as illegitimate access, deceit, and falsification. According to Gordon and Ford, cybercrime is any criminal activity performed on a computer, hardware resource, or network. The Council of Europe's Convention on Cybercrime classifies criminal acts into four classes: 1) Breaches of data, secrecy, integrity, and hardware resources 2) Computer-centered crimes 3) Content-related crimes; and 4) copyright-related crimes. However, these classifications are over the line for some parameters. According to another classification given by Saini *et al.*[41], cybercrimes are categorized as data crimes, network crimes, access crimes, and content-related crimes. Data crimes consist of data-stealing, data interception, and data modification. Network crimes include unwanted interference in the functioning of computer networks to breach data transmitted over the network. Content-related crimes include infringement of ownership and spontaneous cyber hazards. Another explanation of cybercrime is demonstrated by Zhang *et al.*[42], according to which all crimes in which machines or networks are used as aids or targets, the place of crime, and any conventional crime executed with computer resources are addressed as cybercrime. Generally, ICT boosts the rate as well as the domain of criminal actions. The location of crime acts as a catalyst for criminal activities [43]. Internet is also a large platform for criminal acts as was not initially deployed with highly secure protocols. As IoT systems are implemented on the ceiling of the present internet framework; so, the associated cybercrime issues remain unresolved. At last, a large base of the cyber framework enhances the scope not to reveal criminal acts to the public as the criminal acts are executed using virtual methods.

## 3 Existing work on IoT Security and Cybercrime

In the last few years, several surveys have been conducted to impress upon the improvements and research carried out in the IoT systems. In these survey papers, the focus is on the fundamental aspects of IoT. Along with IoT, security issues are also discussed in some of these surveys. There are few dedicated survey papers on IoT security and privacy contention. In the surveys published in the years 2010-2020, Atzori *et al.*[44] discussed the security and privacy aspects of IoT. In the field of security, the main attention is given to authentication and data integrity, and the scope of research is discussed. In the privacy aspect, the authors suggested limiting access to personal data. However, this survey highlights incomplete facts regarding security challenges in IoT. Miorandi *et al.*[45] assumed the implementation of IoT at three fundamental levels, i.e., communication, identification, and interaction. Although the authors highlighted the possibility of many security challenges in IoT but proposed the research on three main issues: privacy of users, data secrecy, and trust. Many burning issues related to IoT security like access control, data integrity, and authentication of the user are not discussed in detail [46]. Gubbi *et al.*[47] discussed security and privacy in the contexts of user identification and authentication, data integrity, and privacy in general. The authors introduced the cloud based IoT paradigm. On the same grounds, few technologies are introduced along with the domains of application of each technology.

In [48], Aggarwal *et al.* discussed security prospectus exclusively from a privacy perspective, whereas other security challenges in IoT platforms are not discussed. Said [49] discussed various IoT architectures along with research issues. In this survey, only challenges faced in physical security and privacy are explored. Moreover, security issues are discussed without giving any viable solutions. Perera *et al.* [50] elaborated that security and privacy challenges are handled at the middleware level in the IoT framework and different layers. In this survey, security is expressed as a normal issue and the authors did not pay any special attention to the research in the field. Granjal *et al.* [51] presented an in-depth review of different security mechanisms and protocols of the time for communication among smart devices. The authors also highlighted the available scope of research. However, on the negative side, the authors did not consider all security standards in their survey but focused on a few only. Sicari *et al.* [52] reviewed security from three different angles: security requirements, privacy, and trust. Under security requirements, the authors explored the issues related to access control, confidentiality, and authentication. The biggest drawback of this work is the inadequacy of the categorization of research activities in the IoT security paradigm. Abomhara and Koein [53] reviewed the security threats along with the security and privacy research challenges in their paper. They stressed research issues like interoperability of diverse IoT devices and authorization.

Mahmoud *et al.* [54] surveyed IoT security principles. The authors also presented various security issues along with corrective measures. The need for advanced technologies to tackle hardware, software, user identification, and wireless communication issues is also discussed. Pescatore and Shpantzer [55] presented the viewpoint of people actively involved in the research of IoT security issues along with the future prospects in the field. They also highlighted that IoT developers should focus more on security issues instead of other ICT systems. Gil *et al.* [56] reviewed various technologies and security models in the context of data-related challenges. The authors impressed upon the collaboration of social networks and IoT and introduced a new concept of the Social Internet of Things (SIoT). IoT security is discussed but the concept of cybersecurity in IoT is not touched. Muhammad *et al.* [57] discussed the various possible attacks in IoT systems. The authors also highlighted the security and privacy challenges faced in the IoT environment by the various sensor nodes. In this survey, the requirements of secure end-to-end communication among smart devices using efficient encryption and authentication methods are suggested. Vignesh and Samydurari [58] reviewed three layers architecture of IoT comprising of application, network and perception layer, along with different types of security threats at these layers. They explained the effect of wireless signals, movement of IoT in the external environment and the dynamism of the network model as the major challenges at the perception layer. At the network layer, the major highlighted challenges are DoS and Man-in-the-Middle attacks. The major issue that persists at the application layer is the variety of application policies.

Razzaq [59] surveyed the different security requirements of an IoT system. He categorized the various IoT attacks into four classes: low level, medium level, high level and extremely high level, and suggested the possible way-outs to handle these attacks. Maple [60] discussed the role of IoT devices in various domains like autonomous vehicles, health, industry 4.0, logistics, smart grid, agriculture, home, offices and entertainment. Along with the security, threats in all these application areas are also reviewed. They highlighted the security issues related to the physical limitations of the things, the versatility of the devices, authentication, authorization and implementation. Various issues related to the privacy of the users are also discussed in this survey. Rughani [61] presented the various challenges faced by crime investigators to collect pieces of evidence from the smart IoT devices available at crime scenes. The author impressed upon the need for corrective measures for the issues to help in crime investigation and make the process easy. Corser *et al.* [62] discussed that to make the IoT systems more secure, the security of smart devices and networks needs to be improved. To improve device-level security, protection of data and dynamic testing plays a major role. To make communication networks more reliable; there is a requirement of authentication, secure protocols, network division and organization. Burhan *et al.* [63] presented a detailed survey on the different layers of IoT architecture along with the potential attacks at each layer. The authors also reviewed various available mechanisms to handle these attacks and their limitations. Security issues in various IoT technologies like sensors, ZigBee, Bluetooth, RFID, Wi-Fi, and 5G networks are discussed in detail.

Noor and Hassan [64] presented the primary objectives of IoT system security. The authors highlighted that privacy of the user and security of the data and infrastructure are the main challenges in the IoT environment. The authors also reviewed various tools and simulators to implement IoT security mechanisms. MacDermott *et al.* [65] highlighted the sharp increase in the usage of digital forensics for crime investigation. The authors also highlighted that the reason for this rise is the increase in smart devices. To cope up with this change, there is a need for regular development in the techniques used for crime investigation. The authors also reviewed various forensic handling methodologies. Sfar *et al.* [66] presented three different aspects i.e., privacy, trust and identification/authentication of IoT security. Under these three aspects, various open research issues like standardization of security mechanisms, reduction in the amount of data transmitted among smart devices, implementation of trust mechanisms to safeguard users and services, implementation of global identification mechanism for things and automatic discovery of devices in the IoT environment are highlighted.

Neshenko *et al.* [67] presented an exhaustive survey on IoT vulnerabilities. The need for endorsement of different advanced technologies like blockchain, deep learning, and cloud paradigms is stressed in IoT security implementation. Various research aspects highlighted in the survey are the requirement of global device identification mechanisms, the need for more security-centric awareness among IoT users, the requirement of more mature security protocols and the adoption of secure IoT application development processes. Zhou *et al.* [68] reviewed four main features of IoT: interdependence, diversity, constraint and myriad. Consequently, the open research issues for these have also been discussed. It is spotlighted in the survey that in IoT systems, the devices are interdependent, so focusing on security mechanisms by considering each device as a standalone will not provide a secure IoT environment. Detection of viruses in IoT devices is also highlighted as an open research challenge in this survey. The issue of sensitivity of the user's personal information is also an area of major concern for academicians and researchers. Lu and Xu [17] elaborated that the privacy and security of IoT systems is the biggest research challenge. The authors presented a detailed review of the state-of-art research going on in cybersecurity. IoT architecture for cybersecurity is discussed in detail. At last, the major research challenges of the domain are also presented. Aydos *et al.* [69] classified IoT vulnerabilities depending upon the types of attack in four different layers: physical layer, network layer, data processing layer and application layer. Depending on these vulnerabilities, the authors proposed a risk-based security model to evaluate each discussed layer of the IoT architecture. Nasiri *et al.* [70] surveyed the security needs of an IoT-dependent health care system. They classified it into two categories: cybersecurity and cyber resilience. Under Cybersecurity, the various features of confidentiality, integrity, availability, identification, and authentication, authorization, privacy, accountability, non-repudiation, auditing and data freshness are elaborated. Under cyber resilience, safety, survival, performance, reliability, maintenance and information security are discussed in detail. Tabassum *et al.* [71] reviewed various IoT security challenges. The authors also demonstrated the role of IoT in industry. In this study, it was presented that how the security issues of individual devices/things used at each layer in the IoT architecture can affect the security of an IoT system.

Servida and Casey [20] presented a detailed study of the vulnerabilities of smart devices. The authors discussed how these vulnerabilities can cause these devices to become victims of attacks. On the positive side, it is featured that these vulnerabilities can help the investigators to fetch digital traces and investigate the crime. Therefore, device vulnerabilities are both challenges and opportunities in crime. Blythe *et al.* [72] highlighted that the IoT environment lacks security features as the devices are not manufactured by considering the security challenges. It is also discussed that at some events even users do not use the available security features of the devices due to a lack of knowledge about the customization of these features. In this work, the authors impressed on the need for standardization of communication and security protocols in IoT systems and highlighted the need for government intervention to assure security at the device level. Adesola *et al.* [73] suggested a novel IoT and Big Data-based smart model to investigate and control criminal activities in Nigeria. The authors also developed a prototype for the model. This model is useful to keep the record of criminals. Abdullah *et al.* [18] discussed the security aspects of IoT by focusing on cybersecurity. Open research issues related to cybersecurity are highlighted along with possible corrective measures. The authors also applied the usage of blockchain technology to strengthen the cybersecurity aspect of IoT. Butun *et al.* [1] presented an in-depth review of various types of security attacks in wireless sensor networks and IoT systems. Various mechanisms for the prevention and detection of these attacks are also discussed in detail. The authors categorized the IoT attacks as active and passive attacks. It is also spotlighted that passive attacks cannot be identified using any mechanism. On the other hand, active attacks violate the integrity and confidentiality of data. Active attacks also cause unauthorized access to user data.

Stoyanova *et al.* [74] surveyed the various available models for digital forensics. Special consideration is given to the methods which are used to extract digital data by maintaining the privacy of the users. The authors presented open research challenges in the field of digital forensics by paying special attention to the need for more advanced forensic analyzing techniques and universally acceptable protocols. Tawalbeh *et al.* [75] discussed the various security and privacy challenges of IoT. The authors also proposed and evaluated a cloud-based IoT security solution. Atlam [76] reviewed IoT architecture and communication technologies. Various IoT security challenges and threats are also discussed. The authors also explained the role of digital forensics in crime investigation. The need for employing real-time techniques in IoT forensics is highlighted as the need of the hour. Al-Khater *et al.*[77] presented a detailed review of various categories of cybercrime in detail. Various cybercrime detection techniques using statistical methods, neural networks, machine learning, deep learning, fuzzy logic, data mining, computer vision, biometric and forensics are also discussed. The authors proposed the requirement of cybercriminal profiling, which can be used as a data set by the investigators in the process of investigation. Table 3 presents the comparison of existing security parameters and approaches in IoT cybercrime.

**Table 3 - Comparison of existing security parameters/approaches/models in IoT cybercrime**

| Author (Year)+++ | Ideology | Parameters | Advantages | Security Issues Discussed |
|---|---|---|---|---|
| Atzori *et al.* (2010) [44] | A survey of Internet of Things (IoT) | Applications, Service Management, Logistics | Applications of IoT were discussed in a detailed manner. | ✘ |
| Miorandi *et al.* (2012) [45] | A survey of Applications and issues of security in IoT | Applications, Security issues of IoT, Research Challenges | Research challenges and issues of securities were explained in detail. | ✓ |
| Gubbi *et al.* (2013) [47] | Application and cloud computing-oriented survey of IoT | Applications, Addressing Schemes, Cloud computing. | Cloud computing and its applications in IoT were discussed. | ✓ |
| Aggarwal *et al.* (2013) [48] | A survey of applications, data management, and research challenges in IoT | Applications, Data management and Analytics, Security, Privacy | Data management in IoT and applications were discussed in detail. | ✓ |
| Said *et al.* (2013) [49] | Evaluation of different IoT architectures is presented. | Hierarchical Architecture, Distributed Architecture | Different IoT architectures were discussed in detail. | ✘ |
| Perera *et al.* (2013) [50] | The authors presented context-aware computing for IoT devices | Context Reasoning, Context Modelling, Context Distribution | Different contexts related to the IoT were presented. | ✘ |
| Granja l*et al.* (2015 ) [51] | Communication Protocols and security parameters are discussed in detail. | Different protocols of IoT communications, Application layer, physical layer, and MAC layer security | The security of different layers in IoT communications was discussed in detail. | ✓ |
| Sicari *et al.* (2015) [52] | Research challenges and existing solutions in IoT security are presented in the survey. | Mobile security in IoT, Trust, and Privacy in IoT, Enforcement in IoT, Authentication, Confidentiality, and Access control in IoT. | Security of IoT was discussed referring to ongoing projects on securing the IoT. | ✓ |

| Author (Year) | Ideology | Parameters | Advantages | Security Issues Discussed |
|---|---|---|---|---|
| Abomhara and Koein (2015) [53] | Research directions concerning IoT security and privacy are presented. | Different cyber-attacks in IoT, security and privacy challenges in IoT, security threats and challenges | Threats to IoT security were discussed in detail. | ✓ |
| Mahmoud *et al.* (2015) [54] | IoT layer architecture and security features are the key aspects of this paper. | IoT architecture, IoT security issues, IoT security countermeasures | Basic architectures of IoT and security issues were discussed in detail. | ✓ |
| Pescatore and Shpantzer (2016) [55] | Surveyed perceptions on IoT, IoT applications, and industry representation by IoT. | Applications, Threats to IoT, Risk management in IoT, Data monitoring | The survey conducted with participants was discussed concerning different parameters related to IoT | ✓ |
| Gil *et al.* (2016) [56] | A general survey of IoT and Context-aware of IoT. | IoT Applications domain, Services for IoT, Data mining for IoT, | Services and data as services were discussed. Applications of IoT are also presented | ✗ |
| Muhammad *et al.* (2016) [78] | A review of security solutions against threats on IoT devices. | Different techniques of attacks on IoT, security and privacy requirements, security solutions in IoT | The threats to IoT security and its measures to counter the threats were explained in detail. | ✓ |
| Vignesh and Samydurari (2017) [58] | A survey on IoT layer architecture and security threats on each layer. | Security features of IoT, Architecture, Security remedies | Security issues in IoT and future directions regarding 5G were discussed in the paper. | ✓ |
| Razzaq *et al.* (2017) [59] | A survey on different types of threats in IoT and their solutions is discussed. | Applications of IoT, Threats to IoT, Analysis of different types of attacks. | Applications of IoT and analysis of different types of security threats were done. | ✓ |
| Maple (2017) [60] | A survey of applications of IoT, Authentication, and identity management of IoT, Security issues of IoT in different applications. | Applications of IoT in automobiles, Health, industry 4.0, agriculture, entertainment, and media. | Security issues on various applications like health and in Automobiles were discussed along with privacy challenges in detail. | ✓ |

| Author (Year) | Ideology | Parameters | Advantages | Security Issues Discussed |
|---|---|---|---|---|
| Rughani (2017) [61] | The authors discussed the architecture of IoT devices along with the security aspects and their application in forensics. | IoT architecture, IoT security issues, and digital security in IoT | Discussions on forensics in IoT and security issues were discussed. | ✓ |
| Corser *et al.* (2017) [62] | The authors laid prime emphasis on the security of IoT devices. | IoT hardware security, Dynamic testing, securing IoT networks | Certain security issues from a hardware and software perspective were discussed in detail. | ✓ |
| Burhan *et al.* (2018) [63] | Compared the different application domains of IoT and discussed the key elements of IoT. | Applications of IoT, Different architecture layers of IoT. | Identity management framework, Security mechanisms for IoT, and improved layered architecture for IoT were discussed. | ✓ |
| Noor and Hassan (2018) [64] | A survey on IoT security, possible attacks on IoT architecture layers are presented. | IoT security attacks on layer review on IoT authentication, trust management, and secure routing. | Attacks on the IoT architecture layer were explained in detail. Secure routing was presented with key features. | ✓ |
| MacDermott *et al.* (2018) [65] | The authors discussed the IoT, the possible crime using, or in IoT devices. | Forensic handling regarding IoT and Crime using IoT devices. | Forensic evidence handling in the smart city was discussed in detail. | ✘ |
| Sfar*et al.* (2018) [66] | A survey on IoT security including discussion on smart manufacturing. | A cognitive approach for IoT, recent research in data privacy Trust management system. | Cognitive and systemic security along with adaptive and context-aware security was discussed in detail. | ✓ |
| Neshenko *et al.* (2018) [67] | A survey on the exploitation of different IoT devices | IoT architecture Security in IoT, IoT vulnerabilities at different architectural layers. | The security aspects of the layer-wise architecture of IoT devices were discussed in detail. | ✓ |

| Author (Year) | Ideology | Parameters | Advantages | Security Issues Discussed |
|---|---|---|---|---|
| Zhou *et al.* (2018) [68] | A survey on security features and privacy in IoT | Attacks on IoT, Threats, and challenges in IoT devices. | Threats to IoT hardware devices were discussed in detail. | ✓ |
| Lu and Xu (2018) [17] | A survey article on IoT cyber-attacks and security schemes. | Different cyber-attack and Layer wise security schemes. | Security schemes for different layered architectures were explained in detail. | ✓ |
| Aydos *et al.* (2019) [69] | A survey of risk and threat assessment on different architecture layers of IoT. | IoT applications, platforms for IoT, IoT protocols, Security, threats, and vulnerabilities in IoT. | Attack on a different layer in IoT was presented along with a risk-based layered approach for IoT security assessment in detail. | ✓ |
| Nasiri *et al.* (2019) [70] | An article on healthcare-based secure IoT environment. | Security requirements in IoT | Cybersecurity requirements were discussed. | ✘ |
| Tabassum *et al.* (2019) [71] | An article on various security issues in IoT. | IoT security requirements Architecture of IoT. | Security issues on perception, application, and network layer were discussed in detail. | ✓ |
| Servida and Casey (2019) [20] | An article on IoT forensics and detection of traces in IoT. | Digital forensics, Privacy, and IoT forensics | IoT forensics and detection, extraction, and parsing of traces from IoT devices were discussed in detail. | ✘ |
| Blythe (2019) [72] | An article on cyber hygiene advice for IoT devices | Security features of IoT devices, Design code of practice for IoT devices. | Standardization of security protocols was the main emphasis. | ✓ |
| Adesola *et al.* (2019) [73] | An article on crime management with IoT-based architecture. | IoT architecture, data collection, and framework for IoT devices. | A crime prediction and monitoring model was proposed. | ✘ |
| Abdullah *et al.* (2019) [18] | A review of cybersecurity issues and challenges. | Cyberattacks, cybersecurity, IoT architecture, and security techniques. | Security techniques at different layers are discussed in detail. The blockchain is implemented to secure the IoT network. | ✓ |

| Author (Year) | Ideology | Parameters | Advantages | Security Issues Discussed |
|---|---|---|---|---|
| Butun *et al.* (2019) [1] | A survey on different kinds of attacks and their countermeasures in IoT devices. | IoT applications, Security attacks on IoT devices, Attacks on different layers of IoT architecture. | Defence against different passive and active attacks on different layers of IoT architecture was discussed in detail. | ✓ |
| Stoyanova *et al.* (2020) [74] | A survey on IoT forensics and its challenges. | IoT forensics components, IoT attacks, IoT security, IoT protocols, IoT layered architecture. | IoT forensics challenges and their solution, secure cloud service models were discussed in detail. | ✓ |
| Tawalbeh *et al.* (2020) [75] | An article on security and privacy in IoT devices. | Generic IoT layers and proposed system model for secure IoT devices. | A system model was proposed using the cloud edge nodes and IoT nodes. | ✓ |
| Atlam *et al.* (2020) [76] | An article on cybercrime, security, and digital forensics for IoT devices. | IoT applications, IoT architecture, characteristics, and communication technologies in IoT, Security threats in IoT. | The security solution of four-layered IoT architecture was discussed in detail. | ✓ |

In this review, we examine the various aspects of IoT systems like architecture, protocols and technologies deployed at various layers and application domains. Potential risks and possible attacks on each layer of the IoT architecture are also discussed. We also present the various security mechanisms and their layers of implementation. Special attention is given to IoT forensics in cybercrime investigations[79, 80]. Various domains like smart homes, smart cities, automated transport, drones, healthcare, etc. are examined to assist cybercrime investigation[81]. The role of various advanced technologies in the investigation of cybercrime is also presented. At the end of the paper, various open research challenges in an IoT environment to contribute towards the process of IoT forensic to aid the process of cybercrime investigation are presented.

# 4 IoT Framework and Applications

IoT is a broad network of devices connected over the internet. It has expanded very briskly in the last few years. Currently, IoT has evolved as a contemporary styled network that acts as an agent to link the real and virtual world. Application domains of IoT are expanding day by day growing from the need for smartphones to different IoT devices like cameras, music players, smartwatches, smart TV, smart VR as shown in Figure 5; So, the probability of cyber-attacks. The fundamental characteristic of IoT applications is to gather data from smart devices and communicate over networks [82]. A gigantic volume of personalized data is gathered by various IoT applications including smart agriculture, healthcare, smart homes, meetings, etc. [83]. This large amount of data is communicated in IoT systems, interpreted and analyzed. In the research carried out by Cisco, there is an estimate of 50 billion smart devices to be plugged into the internet in the current year. It is also predicted that because of advanced features, smart devices will become an important part of day-to-day life in the current year [69]. It is being forethought that the trend of using IoT systems will spike and will keep growing afterward. Due to the vast usage of IoT collected data, a new trend has started. Even data collected on smart devices in an IoT environment can be shared for usage in other real-life applications. However, the biggest challenge in the collected data is the versatility of smart devices supported in the IoT system architecture.

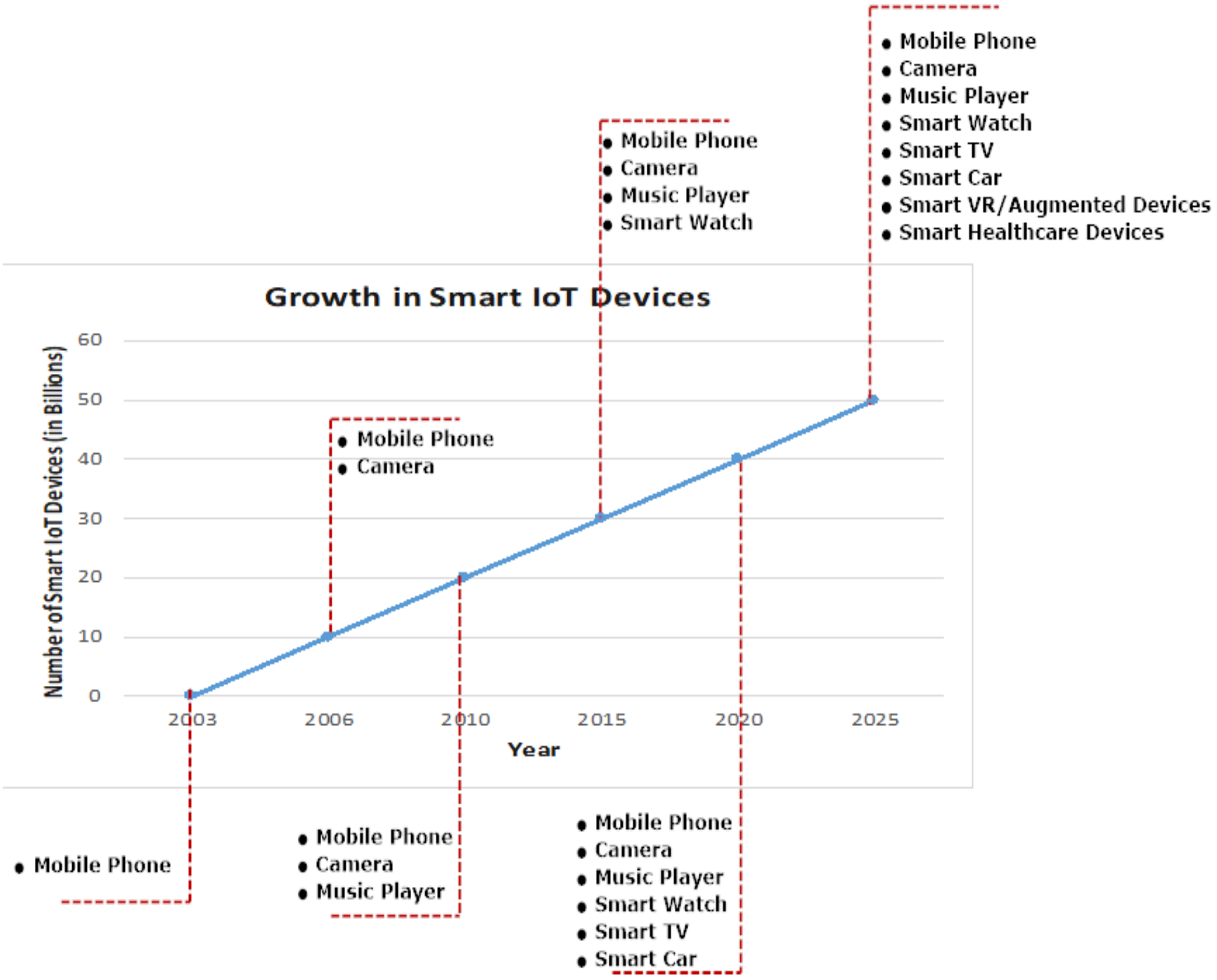

**Figure 5- Growth expectations in the type of IoT devices [69, 84-89]**

### *4.1 IoT Architecture*

There is a need for open architecture to deploy IoT systems to support diverse categories of smart devices and to administer interfacing among them. Many reviews and research articles are available demonstrating the IoT architecture[53]. Fundamentally, IoT systems are deployed on four-layer architecture as shown in Figure 6. These four layers are the application layer, network layer, perception layer and transport layer. This is the basic IoT architecture model which can be practiced with different IoT applications. For each layer of IoT architecture, the possible attacks and the affected domain due to the attack are shown in Figure 6. These technologies help in the process of data collection, interpretation, analysis and communication [90]. Different layers of the IoT architecture are characterized as follows:

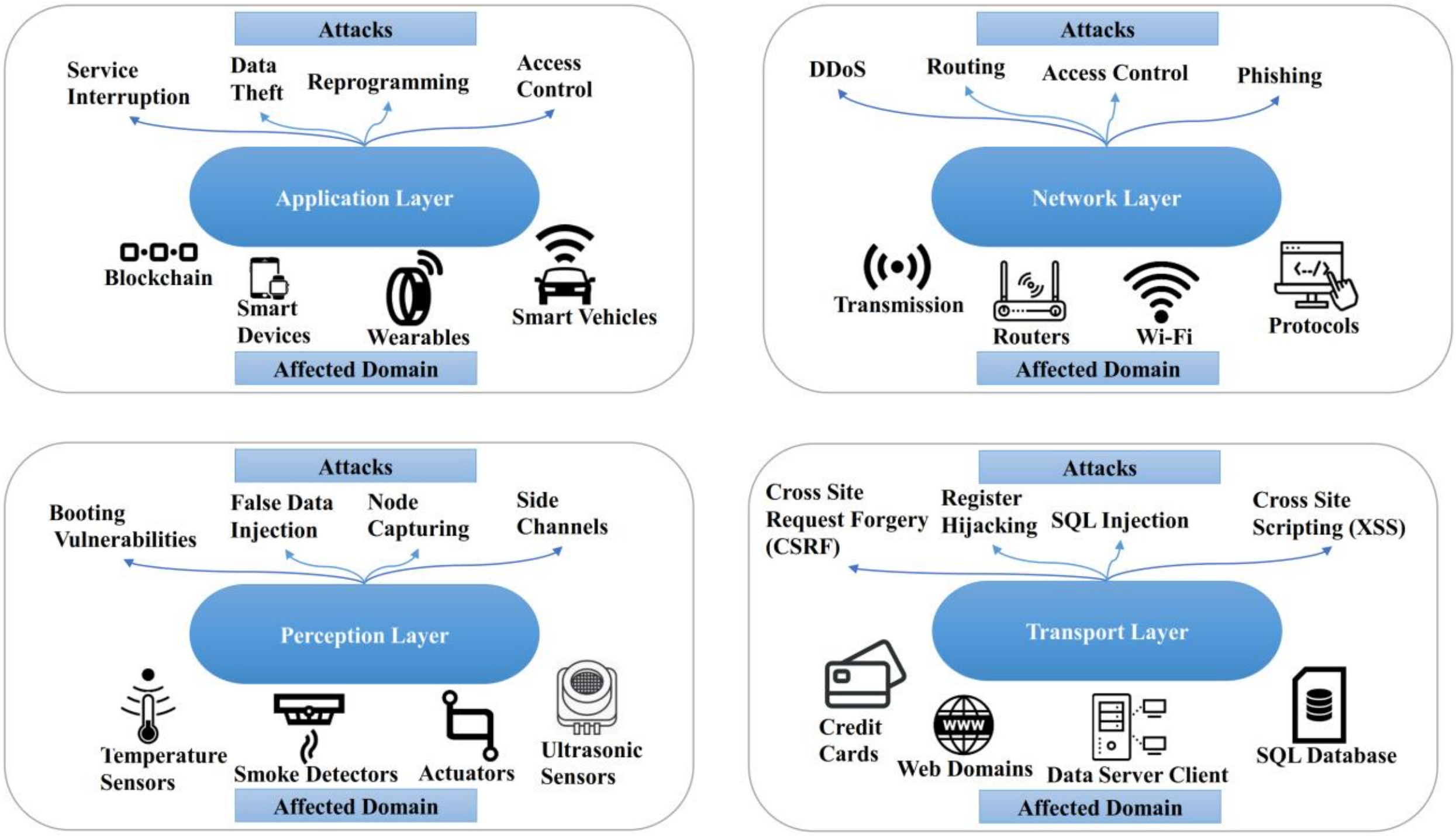

**Figure 6 - Layers of IoT with technologies deployed and possible attacks [25, 91-96]**

- **Perception Layer.** In this layer, data are generated by various smart devices. Data is also gathered by these devices, which can be further communicated within the IoT environment or even to outside applications. This layer works with two types of things: IoT devices and IoT hub nodes [97]. IoT devices identify themselves in the IoT system, whereas IoT hub nodes work as gateways. The data collected through devices are transmitted through gateways [98].
- **Network Layer.** In this layer, communication among IoT devices and applications is managed. The mode of communication may be wired or wireless. Various network security protocols are deployed in the network layer. The IoT gateways are set up at this layer. This layer receives the data coming from the lower layer and maps to the format required by the applications running in the upper layer [99].
- **Application Layer.** The application layer is also interpreted as the service layer. Here, the data gathered by various devices are used, analyzed, interpreted, and presented. This layer can be customized under different policies depending upon the service administered [100].
- **Transport Layer.** The transport layer is responsible for end-to-end communication over the network. It also provides reliability multiplexing along with flow control. Congestion control is also performed in the transport layer [101].

### *4.2 Protocols*

Functionalities provided by the various layers of the IoT architecture are administered by the different protocols deployed in the different layers [102]. Various protocols used at the different layers of the IoT architecture like the Application layer, Perception layer, Network layer and Transport layer are shown in Figure 7. Various protocols deployed in the perception layer are IEEE 802.11 series, 802.15 series, Wireless HART (Highway Addressable Remote Transducer) etc. [100]. The IEEE 802.15.4 is used for data exchange in a long-range wireless personal area network (LR-WPAN). ZigBee and Wireless HARTS are also deployed in the IoT perception layer [103].

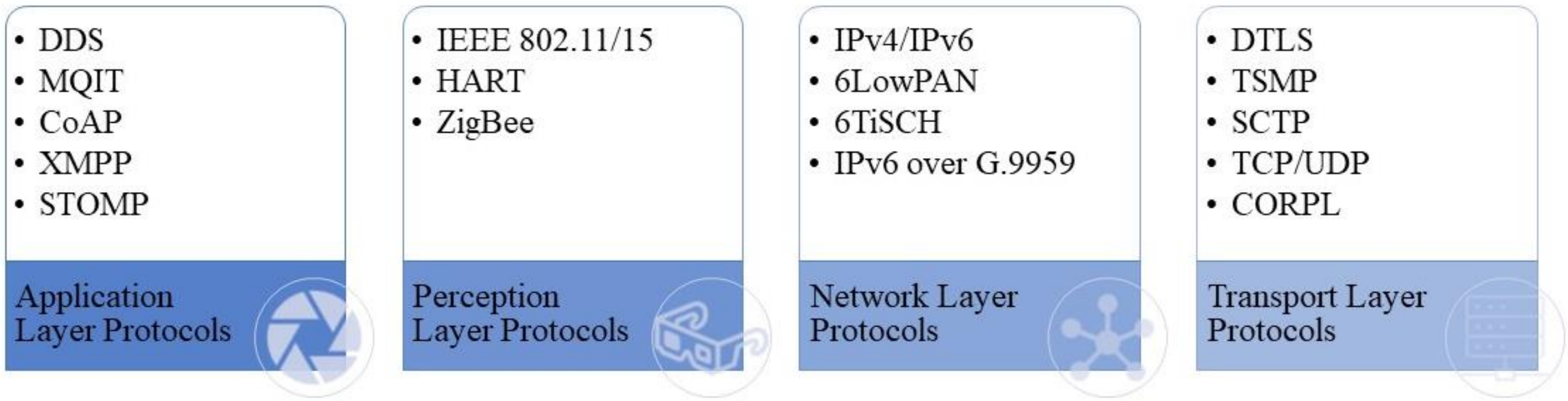

**Figure 7 - Protocols of different layers in IoT [90, 104-106]**

Protocols used in the network layer of IoT architecture are IPv6/IPv4, 6LoWPAN (IPv6 over Low-Power Wireless Personal Area Networks), 6TiSCH (Time-Slotted Channel Hopping) developed by IETF which is IPv6 standard for 802.15.4 MAC layer protocols [104]. IPv6 G.9959is an IPv6 addressing standard for G.9959 MAC layer protocol which was designed for low power devices in a personal area network (PAN). For real-time systems, Data Distribution Service (DDS) is used. This protocol does not require any networking middleware and network programming, which allows the publisher to release specific information. The lightweight messaging protocol used in the application layer is MQTT (Message Queuing Telemetry Transport), and it uses machine-to-machine communication based on TCP-IP. The protocols specially designed for IoT environments e.g. CoAP (Constrained Application Protocol) are used in the application layer for limited hardware. The hardware which does not support HTTP can use the CoAP protocol. The XML-based protocol used in the application layer is known as Extensible Messaging and Presence Protocol (XMPP). XMPP is used for real-time instant messaging and multiparty chat. Simple or Streaming Text Oriented Messaging Protocol (STOMP) is a protocol for message-oriented middleware. It was designed to establish communication between clients and brokers. [107], [108-112]. In the transport layer, Datagram Transport Layer Security (DTLS) is designed to prevent message forgery and tampering. The protocol similar to the time-division multiplexing in the transport layer is Time Synchronized Mesh Protocol (TSMP). It was developed for inter-sensor communication in timeslots. The message-oriented transport layer protocol is the Stream Control Transmission Protocol (SCTP), which uses congestion control to transfer data over a network. For large packets and data, Transmission Control Protocol (TCP) is used in the transport layer in IoT. User Datagram Protocol (UDP) is a protocol for lesser data and is used to send data to the server and suitable for wireless sensor network communication. The extension of the IPv6 routing protocol is Cognitive RPL (CORPL), which was developed especially for cognitive networks. It consists of multiple forwarders with the best node selected to forward the data [113-116].

### *4.3 IoT Application Domains*

The incorporation of smart devices to gather data from our day-to-day life activities make many IoT applications feasible [53]. These applications can be categorized into different domains, summarized below:

- **Personal and Social Domain**. The applications under this domain allow potential users to communicate with the environment or with other users to establish and maintain a social circle [44].
- **Mobility and Transportation Domain**. Under applications falling in this domain, roads and vehicles equipped with sensors and other smart technologies which can gather traffic-related data are covered. This data can help with traffic control and management [117]. Some of the IoT-based transport applications with outstanding performance are Intelligent Traffic Information Service (ITIS) and Traffic Information Grid (TIG) [118].
- **Enterprise and Industrial Domain**. IoT Applications falling in this category include smart banking, manufacturing, logistics, industrial operations etc. [2, 119].
- **Service and Utility Monitoring Domain.** This domain of IoT applications commonly deals with smart agriculture, environment and energy management etc.

### *4.4 Supporting Technologies*

For all applications falling in various IoT domains, different components of the IoT system need to stay connected all time. This is possible only with IoT supporting technologies [53]. The progressive growth of various technologies like sensors, smartphones and software will facilitate different things in the IoT systems to stay connected everywhere and at all times[120]. The fundamental approach to support IoT is to connect the objects in the physical world with the digital world [121]. Numerous technologies device these approaches, as discussed below:

- **Identification Technologies.** The fundamental identification technologies used in IoT are Radio-Frequency Identification (RFID) and Wireless Sensor Networks (WSN). These are used in the perception layer of IoT architecture [44], [31], [119].
- **Network and Communication Technologies.** Both Wired and Wireless technologies (e.g., GSM and UMTS, Wi-Fi, Bluetooth, ZigBee) permit a large number of smart devices and services to be connected [122-124]. Flexible and secure IoT architecture is required for reliable communication among various wireless devices [117].
- **Hardware and Software Technologies.** A lot of research is going on in the field of Nano-electronics to develop vast functionality and economical wireless IoT systems [119]. Smart things with improved inter-node communication will help in the development of smart systems assisting fast application development to support various services in IoT.

### *4.5 Security Challenges*

Every layer of IoT is prone to security attacks and threats. These attacks may fall in any of the categories of active or passive and internal or external attacks [54], [53]. In passive IoT attacks, only the information transmitted on the network is observed but the service is not affected. On the other hand in active attacks, a service stops responding [125]. The various devices and services supported by each layer of IoT are prone to Denial of Service (DoS) attacks. Under DoS attacks, devices, services, and networks become unsalable to unauthorized users. In the same manner, Figure 8 describes the security threats faced by the Perception Layer, Network Layer, Application layer, and Transport layer and services supported at each layer which are discussed below:

- **Security Threats in Perception Layer.** The very first issue faced by the various device nodes functioning in this layer is the intensity of the wireless signals as the signals become weaker due to environmental disturbances. The second issue is related to the physical attacks on the IoT devices as the various IoT nodes usually operate in the outdoor environment. The third issue is related to the dynamic topology of the IoT systems which allows the frequent movement of the IoT nodes in and around the network. Different devices working in this layer use sensors and RFIDs. Because of their limited adequacy from the storage and computational point of view, these devices are prone to different kinds of security threats [53] and [126]. Various kinds of devices operating in this layer are susceptible to Replay Attack, Timing Attacks, Node Capture attacks [57] and DoS attacks. All these security challenges can be dealt with by encryption, access control and authentication [127].
- **Security Threats in Network Layer**. Along with the DoS attacks discussed previously, the network layer of an IoT system can also be targeted for silent monitoring, traffic analysis, and eavesdropping. The major reasons behind these attacks are the remote access and exchange of data. This layer is terrifically prone to a Man-in-the-Middle attack [53]. Eavesdropping is the root cause of the insecure communication channel. Communication technologies and protocols play a major role to stop eavesdropping and further stopping identity theft. As the heterogeneity of devices is a major issue in the IoT systems, it is the biggest challenge to have more secure protocols in the network layer to deal with this diversity. Attackers also misuse the connectivity of the devices to steal user information for future attacks [106]. Along with the security of the network from the attackers, the security of the devices operating in the network is competently important. Consequently, the devices in the network must have the comprehension to safeguard themselves against network attacks. This can be obtained only with secure network protocols as well as smart applications [128].
- **Security Threats in Application Layer**. Lack of standard policies related to IoT systems causes many security challenges in the IoT applications and their development. As a variety of authentication mechanisms are used in different IoT applications, it is difficult to warrant data security and user authentication. The second major challenge is to deal with the interaction of the user with applications, the volume of data exchanged and to manage the different applications. The IoT users must be decked to decide what they wish to share about them and how that information is to be used and by whom [54].

- **Security Threats in Transport Layer.** Common threats in the transport layer include Cross-site scripting (XSS). In this type of attack, the malicious user injects client-side-based scripts like Java, HTML, or VBScript into a webpage that is frequently visited by the user. These scripts will be masked as valid requests between the browser (client-side) and the webserver. It can lead to data theft and manipulation. The other attacks include session hijacking, Cross-site request forgery (CSRF), and Lightweight Directory Access Protocol (LDAP) injection [129].

Table 4 describes the taxonomy of various attacks and defence mechanisms at different layers of IoT devices.

**Table 4- Taxonomy of various attacks and defence mechanism at different Layers.**

| **Application Layer** | | **Network Layer** | | **Perception Layer** | | **Transport layer** | |
|---|---|---|---|---|---|---|---|
| **Attack** | **Possible Defence Mechanism** | **Attack** | **Possible Defence Mechanism** | **Attack** | **Possible Defence Mechanism** | **Attack** | **Possible Defence Mechanism** |
| Common Injection Attack | OpenHab Technology<br><br>IoTOne Technology | Node misbehaviour and Service Attack | Reputation based | Cipher Text Attack | Encryption based on Hash | Flooding Attack | Compressed DTLS Header |
| Attack on Privacy | Preference based Protection | Identity Theft Attack | Identity Management Framework | DDoS Attack | PKI Protocol | Replay Attack | Compressed IPsec |
| Identity Spoofing Attack | Security Framework based on Identity | Fault Injection Attack | Risk based Adaptive Framework | Phishing Attack | Secure Authorization OAuth | Routing Attack | ECC DTLS 6LoWPAN Border router ECC |
| | | Side Channel Attack | SDN enabled IoT | Side Channel Attack | Lightweight Cryptography | | |
| | | No Forwarding Attack | Cooperative Nodes Protocol | Crypt-analysis Attack | Framework based on Embedded Security | | |
| | | Eavesdropping | Cluster based IDS | | | | |

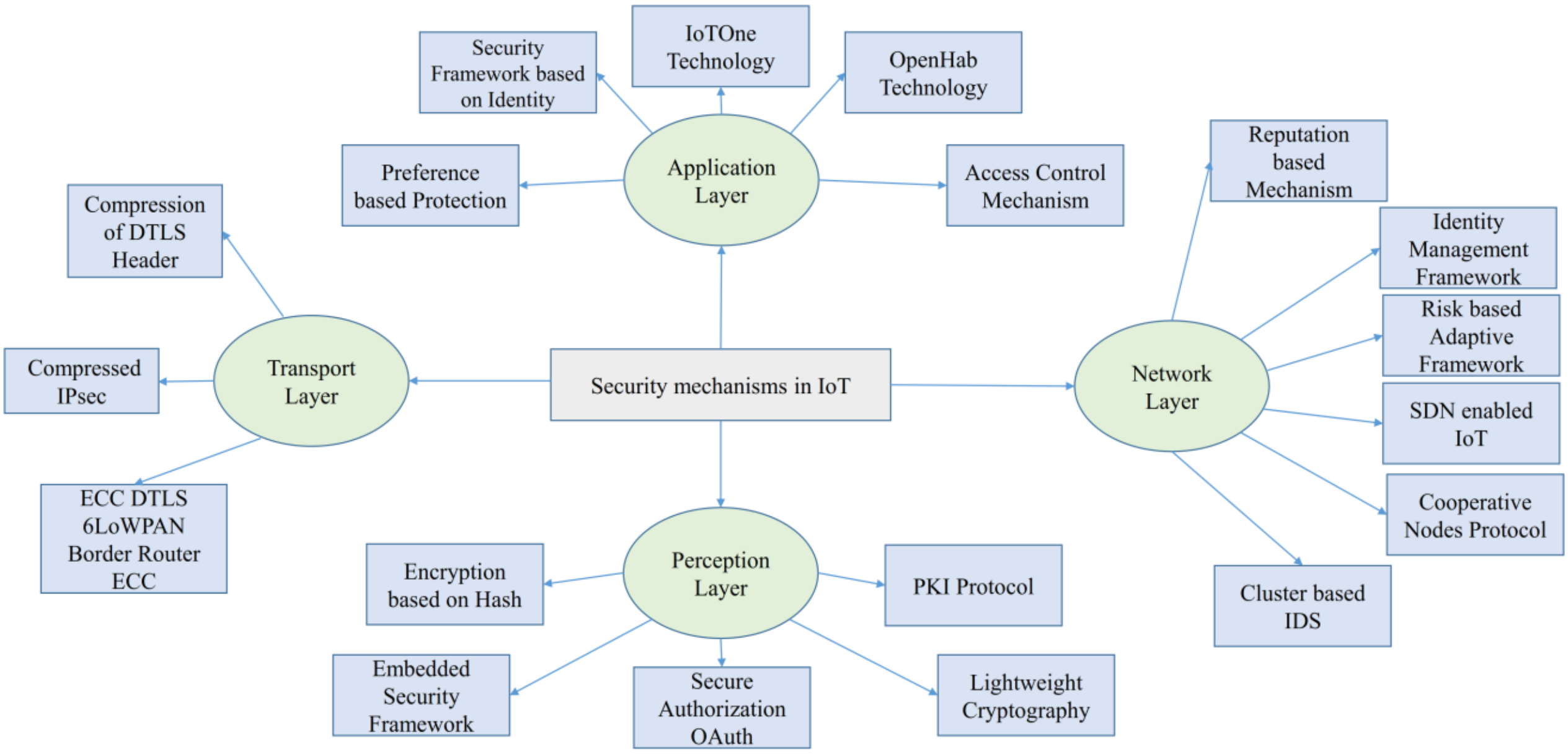

**Figure 8 - Existing security mechanisms for the protection of IoT applications [54, 63, 130-143]**

### *4.6 IoT Security Mechanisms and Measures*

Security is a demanding affair that persists in IoT systems. The benefits of the IoT system cannot be obtained without addressing different security issues [63], [144]. Various security mechanisms proposed by various researchers to safeguard different IoT applications are shown in Figure 8. Different security mechanisms used in the Perception Layer of the IoT systems are Encryption and Hash-based security [130], [145], Public Key Infrastructure (PKI) Like Protocol [131], [146], Secure Authorization Mechanism with OAuth (Open Authorization) [147], [132], Lightweight Cryptographic Algorithms [133], Embedded Security Framework [134], [148]. The Network Layer of IoT is protected with Identity Management Framework [136], Risk-Based Adaptive Framework [137], Association of SDN (Software Defined Networking) with IoT [149], Cooperation of Node-Based Communication Protocol [138], Reputation System-Based Mechanism [139], Cluster-Based Intrusion Detection and Prevention System [140].Various security mechanisms implemented in the Application layer of IoT are Preference-Based Privacy Protection Method [141], [150], Access Control Mechanisms [142], [151], OpenHab Technology, IoTOne Technology [143] and Identity-Based Security [152], [153]. All these security mechanisms about the security provided by the different layers of IoT are compared below in Table 5.

**Table 5- Comparison of existing IoT security mechanisms in different layers**

| Method Name | Layer | Description | Issues focused on the method |
|---|---|---|---|
| Risk-based adaptive framework | Network | Each portion of the four portions performs its tasks and acknowledges the other. | It keeps watching for attacks. It removes the incoming attack at the second portion [137]. |
| Preference-based Privacy protection | Application | The service provider, client, and third-party initiate communication in a secure environment. | Between the client and the service provider, the third party acts as a bridge and keeps a check on the security provided to the client through the service provider [150]. |
| OpenHab in the Application layer | Application | Provision of Security. | The device mismatch is not supported but registration is simple [143]. |
| PKI protocol | Perception | A message is sent by the base station to the destination consisting of a public key. | The message is delivered independently without compromising security [146]. |
| IoTOne | Application | OpenHab technology issues are solved. | A device mismatch is allowed. The request is sent by the client to the server for the verification of the user [143]. |
| Security framework based on Identity | Application | Registration, policy, client, and user authentication are part of this system. | Admin describes the policies. Users and all other resources are managed by the framework based on policies [152]. |
| Encryption based on Hash | Perception | Encryption algorithms and hash functions are used in parallel. | The integrity of the message is checked [145]. |
| Mechanism-based on the secure authorization | Perception | RBAC and ABAC mechanisms and systems are based on client-server. | Resources are provided by the server to the client on request, thus making the system more secure [132]. |
| Lightweight cryptographic algorithms | Perception | Messages are converted by using keys. | Plain text from the message is converted to a cipher using hash functions, symmetric, and asymmetric keys [154]. |
| Embedded Framework of security | Perception | Memory Operating system and run-time environment are secured. | More secure memory management, secondary storage, and run time environment to the users [134]. |
| The Framework of Identity Management | Network | Communication is done via service and identity. | Information about the user is confirmed by the identity module to protect the users from the attackers [136]. |
| SDN with IoT | Network | Low cost and lesser hardware are used for better performance. | IoT agents and controllers are provided security by SDN as all communications are done through SDN [149]. |
| Mechanism-Based on Reputation | Network | Data structures, namely, the reputation table and watchdog mechanism, are maintained by the node to prevent intruders. | Ad hoc communication-based system [139]. |
| Heterogeneous fusion mechanism in IoT | Transport | Prevents disclosure of data and information. | Roaming authentication security in the heterogeneous environment [155]. |

## 5 Role of Digital Forensics in Cybercrime Investigation

Although crime always persisted in society, the aids of committing the crime evolved and grew with time. With the advancement in technology, criminals have come up with new and technology-equipped methods to commit crimes called cybercrime. In the past, criminal inquiries depended on the investigation of the physical evidence and crime locations along with witnesses. However nowadays in the internet era, crime scenes may comprise smart IoT devices, computers, etc. [65]. Consequently, the process of criminal investigation may consist of the analysis of digital evidence [156].

### *5.1 Digital Forensics*

Digital evidence may consist of a variety of elements. Primarily, the evidence would consist of smartphones, laptops, computers, hard drives, USB, etc. As everyone can have any of the above devices, a large volume of data will be available for analysis. But a major hindering factor in the analysis is the variety of formats in which data is available on these different devices [65]. As there is a big change in the type of evidence with time, so there is a need for new techniques to handle this change efficiently. Just like traditional forensics, digital Forensics is a domain that interprets digital data [74]. Digital Forensics experts collect, preserve and analyze digital evidence [157]. Rogers states, "The science of digital forensics has developed, or more correctly is developing; while this science is arguably in its infancy, care must be taken to ensure that we do not lose sight of the goal of the investigation process namely identifying the parties responsible" [65], [158]. During the design and development of new techniques to analyze digital evidence, it is mandatory to consider other aiding domains to develop and support in the process of the criminal investigation. A digital forensics approach deploys a framework for the techniques to be used in digital forensics-dependent investigation [159].

### *5.2 IoT Forensics in Cybercrime Investigation*

The IoT Forensics can be observed as a sub-domain of Digital Forensics. IoT forensics is a comparatively new and less scrutinized area. Its fundamental aim falls in line with Digital Forensics i.e., to collect and analyze digital evidence legally and accurately [74]. In IoT Forensics, data could be collected from sensors, IoT devices, networks, and clouds [160]. IoT Forensics can be categorized as device-level forensics, network forensics, and cloud forensics as shown in Figure 9.

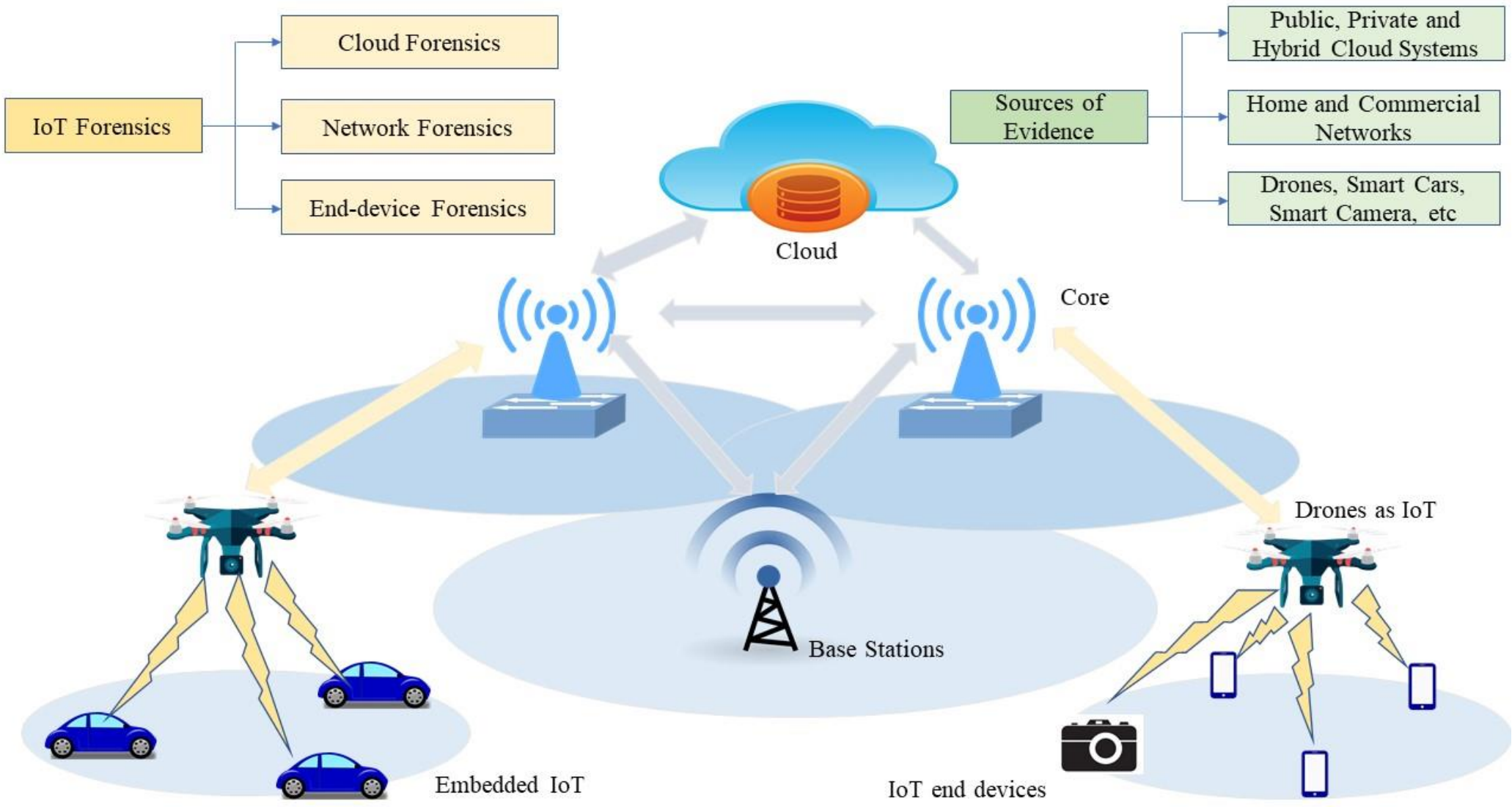

**Figure 9 - IoT forensics components [74, 157, 160-162]**

The basic contrast between Digital Forensics and IoT forensics depends upon the devices examined in crime investigation. In Digital Forensics, the various devices under examination may be computers/laptops, servers, tablets, and smartphones [27]. Although IoT forensics has a wider area of applicability like smart homes, smart vehicles, drones, general IoT systems, etc., the published literature in the area in comparison to Digital Forensics is less.

- **Smart Homes**. It has been observed that during criminal investigation, smart home devices can provide compromising information [163]. Usually, the main components of these devices are microphones and motion detectors. These devices play a major role in identifying the location of suspects. There are three main categories of devices to collect forensics: active, passive and single-malicious active. In [163], two smart devices i.e., light and bulb, have been experimented by the authors. It has been observed that a large amount of data can be collected even with these passive devices, which can help to identify the activity executed in a specific timestamp. The design of another smart home solution, i.e., Forensics Edge Management System (FEMS) is discussed in [164]. The focus of the proposed system is to administer security in smart homes along with forensics assistance. Although a variety of features ranging from automatic detection to intelligence and flexibility are the main, this system is with two limitations i.e. complex implementation and testing. The authors stated in [165] about the security concerns in smart devices. It is impressive that the security threat in an IoT environment increases with the increase in the number of devices in the network. Consequently, the need for IoT forensics arises. In this case study, special attention is given to the IoT forensics in smart homes. The authors also highlighted the need for advanced IoT forensics because of the different IoT challenges. It is expected that in the coming future, smart homes will become widespread. Therefore, a seven-step methodology is proposed for easy investigation in smart home surroundings [166]. It is highlighted by the authors that the proposed framework assists in evidence collection and storage. However, it needs to be tested with a true home automation system.

- **Smart City and Vehicle Automation.** Smart Cities are computerized environments; also termed cyber-physical ecosystems to enhance the utility of traditional city infrastructure like parking spaces, power grids, gas pipes, etc. [74] [167]. In this way, better services can be provided to the residents [168], [169]. One of the important examples i.e. smart parking is the area of the major concern of most of the city administrations and auto-tech companies [170]. The network of smart vehicles assists the exchange of information between the vehicles and the environment [27]. These smart vehicles have aided various important areas like road safety and traffic administration. However, they have also raised many issues in the concern of digital forensics. In a case study [171], a new framework named 'Trust – Internet of Vehicles ( IoV)' is proposed by the authors for dependable investigation. It assists to gather and save dependable evidence from a network of tremendously scattered smart vehicles [172]. This framework is also very useful in preserving the evidence and assuring the integrity of the saved evidence. In [173], various threats to smart vehicles are reviewed by the authors. The authors also proposed and tested a new technique to investigate smart vehicles. However, still this technique needs to be validated with the data produced by a network of smart vehicles in an actual scenario.

- **Drone Forensics.** In [174], the authors proposed a new approach for the forensic analysis of data gathered through drones. The reference data used for forensic analysis were collected from the DJI Phantom III drone. Drone Open-Source Parser (DROP), a new tool to format the data and prepare for internal storage of the system, is also proposed. It is elaborated by the authors that the drone is controlled with the help of mobile and various types of data files are also found on the controlled mobile phone. The data collected in these files aid to identify the location, flight time, and other related information of the drone under observation. However, the main limitation of the work is it focused only on one type of drone; so, the work is needed to be extended for other types too.

- **Cloud Forensics.** Cloud forensics acts as a backbone to IoT forensics. In [175], the authors proposed a new technique to gather and analyze data from the newer BitTorrent Sync peer-to-peer cloud storage service [176]. The data is generated by experimenting with a variety of diverse smart systems. The authors observed that data stored in various log files, installation records and metadata can be recovered. It is highlighted that the state of data in memory should be conserved for accurate forensic analysis. But proposed method has not been legitimized by actual device manufacturers [177].

- **Smartphone Forensics.** In the modern era, people are highly dependent on smartphones. Smartphones play a major role in the exchange of text, audio and video data. Criminals can commit different types of crimes using smartphones like transaction fraud, harassment, child trafficking, pornography etc. It is very difficult to elicit data related to the above activities from smartphones for forensic analysis. To solve this issue, the authors conducted a study [178]. In their study, the Samsung Galaxy S3 phone is used as a device of the experiment for data extraction. It has been observed that to transplant a mobile phone is a tedious activity as it is always associated with the risk i.e. damage to PoP components. Authors in [178] proposed a new methodology named PoP chip-off/TCA. This methodology aids in the transplantation of mobile phones. A new technique was designed and experimented for the successful forensic transplantation of a cryptographic Blackberry 9900 PGP mobile phone.

- **Healthcare Forensics.** The Healthcare sector is one of the most prone domains to major security threats. The main reason for this is the diverse nature of medical applications and the heterogeneity of the types of equipment used; so, a broader surface for attacks [74], [179]. Besides evolution in the health care industry plays a major role in the development of human life, various smart health monitoring systems also put the security of patient's medical data at risk. IoT based fitness systems could be targeted by attackers to steal the data of the users, which can be further misused [180]. Numerous medical identity thefts have been identified in the past which expresses the importance of medical data. In the domain of medical health services and applications, a compound annual growth of 29-30% is expected from the year 2019-2025 [181]. Many fitness wearables can be used as a source of evidence in criminal investigations as these gadgets keep on storing the data related to routine activities of the users at the back end passively. Thus, although these gadgets were designed to maintain the health status of the users but can also be used as digital evidence [22]. The number of users, smart watches, and fitness bands are increasing day by day; so, the study of these IoT devices has become the center of interest for forensics practice. According to the authors in [182], the data extracted from these gadgets may be personal to the users. Therefore, special attention should be given to the security of this data. As the number of security-related issues is increasing exponentially, there is a requirement of more advanced techniques to ensure the security of data [183].

- **General IoT System Forensics.** In [22], the authors came up with a new investigation platform for diverse IoT systems. A risk judgment scheme dependent on STRIDE and DREAD methods was designed and modeled. It was discussed with the help of these two exemplary models that cybercrime committed in the IoT environment can even cause serious risks like death. It was observed by the authors that most of the IoT systems are not deployed with default security measures; so, it possesses high risk. A study was carried out [184] to analyze the significance of the sync data in evidence analysis. Sync data contributes to the fair investigation of the digital witness. A survey was conducted [185] by the authors to study and analyze forensics investigation techniques for data stored in the system memory. Few meaningful alterations to the operating systems were also impressed upon in this study. In [186], data contraction and partially automated analysis techniques to handle a large volume of digital evidence were suggested. This technique assists in the analysis of a variety of IoT data gathered. In [187], the authors discussed the approaches of gathering, saving and communicating digital evidence in a secure way to a genuine destination. Some technologies to bring it into practice were also highlighted by the authors. Along with, the basic components of the electronic evidence were also described.

In [188], a novel approach to club cloud-native and cloud-centric forensic for the Amazon Alexa ecosystem was proposed. A new framework named “Probe-IoT” is presented in [189], which aids to identify criminal evidence in the IoT environment using electronic logs. These logs preserve the complete information regarding all data exchanges between things, users and cloud services. This framework was not tested experimentally but conceptually safeguards the integrity of the evidence. In [190], the authors presented a novel model for IoT forensics named PRoFIT to ensure the implementation of standards during forensic analysis. This model was tested in the true IoT environment deployed in a coffee shop. 1-2-3 zone approach is applied by the authors [191] for IoT forensic analysis. According to the authors, concerned persons and pieces of evidence fall in zone 1, things or devices near to the boundary of the network fall in zone 2 and devices exterior to the network are capped in zone 3. This approach was developed to support accurate IoT investigation. However, the practical implementation of this approach is comparatively challenging. The authors in [192], presented a new framework dependent on three-layer architecture. The proposed framework has many advantages to ensure data security with only one disadvantage that it is not much suitable to cope with the limited resources of IoT devices like processing power, battery, etc. The researchers in [193], proposed the design of a new model to help the forensic expert for IoT evidence analysis. This model was proposed to preserve volatile data in IoT devices. This work was planned as an extension of previous research. Using this model, forensic experts can investigate a broader surface in the data domain. However, it has been observed that this model is laborious to implement in a true environment. In [194], the authors presented IoT forensics in a new way. In this work, the IoT domain was methodically explored to disclose the various challenges in the domain of digital forensics. A novel technique named Forensic Aware IoT (FAIoT) was introduced with a focus to uncover new details in an IoT environment. However, the applicability of the approach is doubted as it was not verified in the IoT environment. The authors [195] analyzed prominent technical issues in digital forensics which can obstacle the identification of important facts for investigation. Various research issues, which can significantly improve the process of digital forensics, were also highlighted. Different types of attacks that are frequently planned on the devices in an IoT environment were discussed in [196] along with the complexity which they add to the digital investigation. The hackers use a large number of random UDP attacks at the same time by using UDP datagrams of varying sizes. Consequently, the denial of service is caused. Authors introduced a novel approach to handle these types of attacks by identifying their originators. However, the true implementation of the proposed work is lying on the back front. A number of patents are granted in the development of digital forensics in past. Table 6 represents the patents granted in recent years. Many applications and digital forensics have been developed to prevent cybercrime. Table 7 presents the list of real-time digital forensics applications that support various operating systems and other platforms to prevent cybercrime in IoT devices.

**Table 6- Various Patents granted in the development of Digital Forensics.**

| Patent Title | Year | | Inventor | Country | Key features |
|---|---|---|---|---|---|
| | Filed | Published | | | |
| **Differencing Engine for Digital Forensics** | 2018 | 2020 | Monsen and Glisson [197] | US | Anomaly detection to mitigate the security attack on cloud-based servers. |
| **Forensic Investigation Tool** | 2017 | 2019 | Jon D. McEachron[198] | US | Digital Investigation tool capable of recovering and decrypting the content. |
| **Forensic system, Forensic Methods, and Forensic Program** | 2015 | 2016 | Morimoto *et al.* [199] | US | A medium to acquire and analyze the digital information in a server or a plurality of computers. |
| **Devices and methods for providing security in a remote digital forensic environment** | 2016 | 2017 | Kang *et al.*[200] | US | A method for collecting digital evidence from the target system. Analysis of the collected evidence to be done at a remote location. |
| **Method and apparatus for Digital Forensics** | 2008 | 2012 | Choi *et al.*[201] | US | A method to perform digital forensics by extracting page files from the target stored medium. Also, extract features from the extracted page file. |
| **Systems and methods for provisioning digital forensics services remotely over public and private networks** | 2012 | 2015 | Shannon and Decker[202] | US | A method to collect and analyse electronically stored information over public and private networks using cloud computing. |
| **Digital Forensics** | 2009 | 2014 | Buchanan *et al.*[203] | US | System call information is acquired from the device under test. The acquired data is converted into a sequence format for further investigation. |
| **Methods for data analysis and digital forensics and systems using the same** | 2011 | 2014 | Gil *et al.* [204] | US | It comprises an online data forensic server to acquire and analyze the usage history of a device. It also issues a timestamp to the collected data. |
| **Forensic digital watermarking with variable orientation and protocols** | 2001 | 2008 | K. Levy [205] | US | A method of forensic digital watermarking on the randomly selected orientation in the content signal. |
| **Secure digital forensics** | 2007 | 2011 | Carpenter and Westerinen [206] | US | To perform an audit of computer processor status and memory, a security module is designed. This can be done using a separate hardware path to access the processor register data through a debug port. |

**Table 7- List of Real-Time Digital Forensics Applications to Prevent Cyber Crime**

| Software | OS/Support | Features | Sources |
|---|---|---|---|
| **E3 Universal** | Window, Linux, macOS, iOS | IoT analysis, Cloud data imaging, and analysis, Registry analysis, Email investigation, JTAG, and chip dump processing | https://paraben.com/digital-forensic-tools-6/ |
| **WireShark** | Windows, Linux, macOS, Solaris | VoIP, GUI, Offline analysis, WAN/LAN analyzer. | https://www.wireshark.org/ |
| **Autopsy** | Windows, Linux, macOS Android | Registry analysis, LNK file analysis, Timeline analysis, File type detection, Email analysis | https://www.sleuthkit.org/autopsy/ |
| **Paladin** | Linux | Device cloning support for many forensic image formats: E01, Ex01, RAW, VHD and AFF, Disk Manager, Automatic logging | https://sumuri.com/software/paladin/ |
| **Dumpzilla** | Unix, Windows | Forensic information extraction from firefox, SeaMonkey browsers including cookies, bookmarks, Web forms, SSL certificates, Browser saved passwords. | https://tools.kali.org/forensics/dumpzilla |
| **SIFT (SANS investigative forensic toolkit)** | Linux | File system support, different Evidence image format support, Rapid scripting and analysis | https://digital-forensics.sans.org/community/downloads |
| **Toolsley** | Web-based | File repairing, text encoding, File identification, File signature verification, Binary inspection, CRC tool | https://www.toolsley.com/ |
| **NetworkMiner** | Windows, Linux, MacOS X, FreeBSD | Live sniffing, OS fingerprinting, Geo IP localization, DNS whitelisting, Audio extraction and playback of VoIP calls, PCAP and PcapNG file parsing | https://sectools.org/tool/networkminer/ |
| **Elcomsoft** | Windows, macOS, iOS | Password recovery, Cloud explorer, Disk decryption, Wireless security auditor | https://www.elcomsoft.co.uk/ |
| **Belkasoft X** | Windows macOS, Linux, iOS, Android, Blackberry | E01/DD imaging, Hash set analysis, Registry viewer, plist viewer, Artifacts viewer, SQLite viewer | https://belkasoft.com/ |

## 6 Advanced IoT Security

Smart devices and applications in the various application areas of IoT make human life more comfortable, but also make IoT systems more vulnerable to cyber-attacks. These devices and applications are connected to the internet, which creates new opportunities for cybercriminals to enter the IoT environment. Cybercriminals can enter an IoT system through routers and can damage it in many ways. Although several security mechanisms are available in IoT, advanced technologies like Artificial Intelligence (AI), Machine Learning (ML), Neural Networks (NN), Blockchain technology, Fog computing and Edge computing are playing a major role to handle cyber-attacks and helping to control cybercrime [207, 208]. Authors in [209], discussed in brief the various kinds of security threats in an IoT environment. The need for a dynamic and quick system to safeguard the IoT systems against cybercrime is impressed upon. The authors proposed a hybrid system to detect cyber-attacks using AI & ML in a cloud computing environment. Both types of attacks i.e. device level and network level can be detected with this model. According to the authors, it is considered by the security experts that AI & ML provides very powerful security mechanisms as even future attacks may be predicted based on past IoT attack data. Consequently, this system does not wait for the occurrence of attacks but can predict in advance. The main limitation of the system is that it can work only with standard data formats for prediction. Along with ML provides solutions to DoS attacks, eavesdropping, spoofing and privacy leakage in an IoT environment [25]. Authors in [210], presented a multilayer architecture to associate the various devices within IoT to make them accessible throughout the network at all times. To deal with the security issues of end nodes and to provide more credible services, a novel framework using NN was proposed. According to this framework, security issues need to be tackled in each layer of the IoT architecture. Each end node configured using this framework will have the potential to self-monitor and recover after any unwanted event/attack. In the proposed framework, a NN based adaptive model was used for the automatic recovery of the nodes. In [211], the authors presented an Artificial Neural Network (ANN) approach to control Distributed Denial of service (DDoS) attacks. The ANN was tested in a simulated IoT environment. The results obtained with the proposed technique were found to be 99.4 % accurate and this technique is capable to identify numerous DDoS/ DoS attacks. Authors in [25], highlighted that the incorporation of blockchain in IoT systems has numerous benefits. The distributed architecture of blockchain reduces the risk of failure of data storage nodes. Thus, it leads to more secure data storage in the IoT environment [212] [213]. The concept of data encryption is used by the blockchain for data storage in the IoT environment; so, there are less chances of storing damaged data in things [214]. The augmentation of blockchain with IoT also helps to prevent unauthorized access, data loss and spoofing attacks [215]. Various challenges in IoT along with the workable solutions administered by the blockchain technology are discussed below in Table 8.

**Table 8- Theoretical solutions offered by deploying Blockchain in IoT Framework to prevent Cyberattacks**

| Challenges in IoT | Specifications | Theoretical Blockchain Solution |
|---|---|---|
| **Defects in Architecture** | A point of failure exists in IoT devices that affect the device and the network. | Validation can be done using blockchain. The data is also verified through cryptography to ensure that a legitimate sender has sent it [216]. |
| **Manipulation of Data** | The data extracted from IoT devices is manipulated and is used inappropriately. | Using blockchain, the IoT devices are interlocked due to which the system rejects any kind of change in data through IoT devices [217, 218]. |
| **Service inefficiency due to heavy load on the cloud server** | Cloud services malfunctions due to cyber-attack, power failures, or bugs in software. | Data records are uploaded on different nodes on the network. Due to the same data in different nodes, there is no single point of failure [219, 220]. |
| **Traffic and cost management** | The handling of the exponential growth in IoT devices is a tedious task. | The IoT devices can be connected and communicated through peers bypassing the central servers through the decentralization feature [221, 222]. |
| **Privacy issues in IoT devices** | The user data present in IoT devices are more vulnerable due to cyber-attacks. | The permissioned blockchain can eradicate this problem [223-225]. |

In [25], the authors discussed that a large volume of data is generated by diverse devices in the IoT environment. It is very taxing to shift the entire data to the cloud for real-time analysis; so, the concept of fog computing evolved. Under this concept, the cloud framework is extended to the edge of the network [226]. Fog computing can handle various IoT security attacks like a man-in-the-middle attack, data transit attacks, eavesdropping and resource constraint issues very efficiently [227]. Therefore, various characteristics and possible solutions deployed by fog computing are shown in Figure 10. Authors in [25], discussed that the edge computing framework is an expansion of cloud computing. The location of the computational power and analysis mechanisms differentiate edge computing from fog computing in an IoT environment [228]. In edge computing, both these potentials reside at the edge [229]. The various devices in the IoT system coordinate to establish a network and perform various computations required for data analysis within that network [230]. Therefore, the need to communicate the data outside the device reduces which contributes to improved data security in the IoT applications. On the same grounds, this framework also aids to minimize the communication cost of data [231]. The concept of edge computing helps to handle data breaches, data compliance issues, safety issues, and bandwidth challenges in an IoT environment [232].

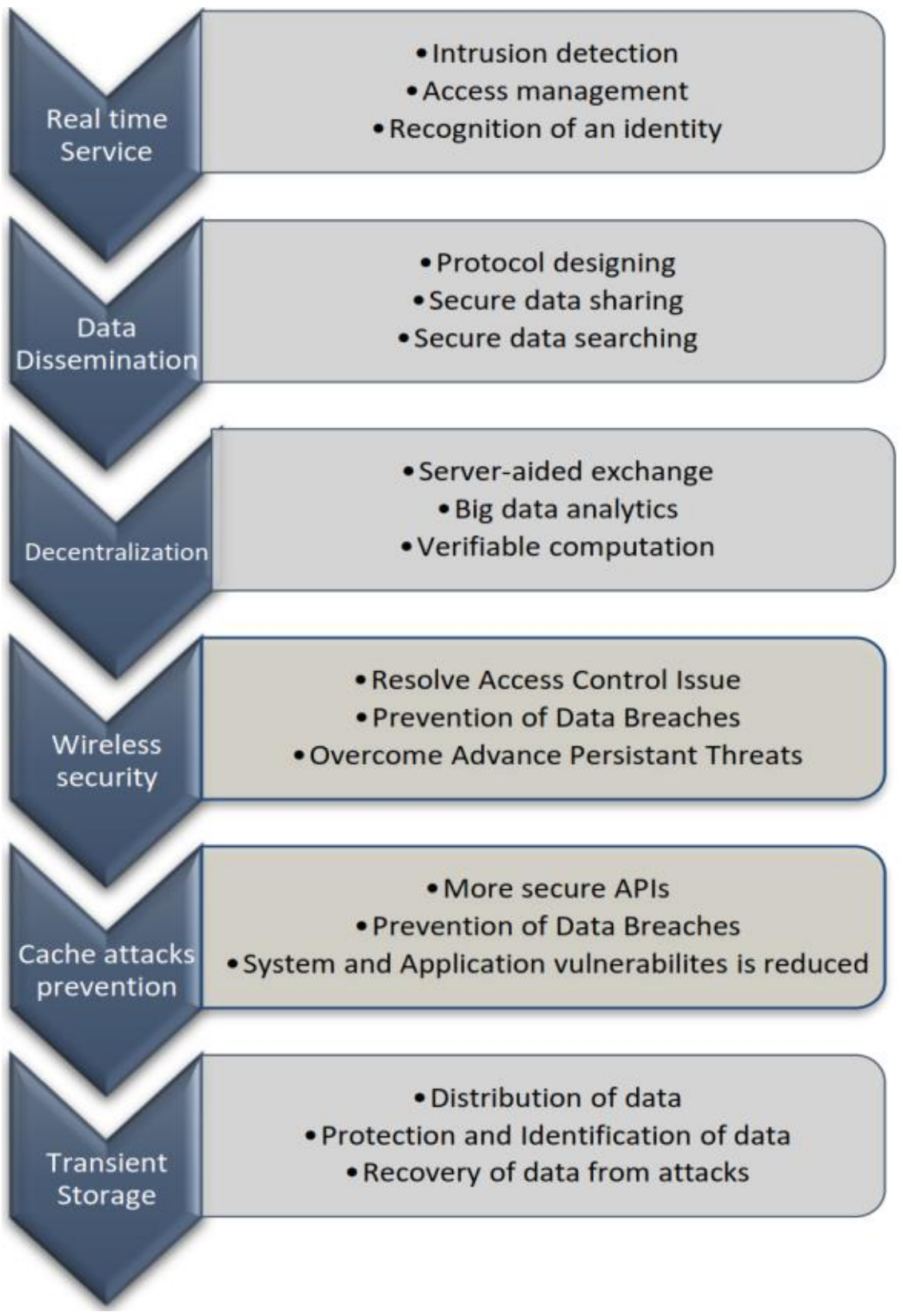

**Figure 10- Possible solutions offered through Fog Computing [233-244]**

# 7 Road map of problems in IoT Forensics

IoT Forensics is a complicated and regularly emerging domain. It plays a very crucial role in cybercrime investigation. However, many challenges need to be addressed very carefully. These challenges open the doors for further research in the field of IoT forensics [74]. Thus, the main objective of this section is to show a path to the researchers in the domain of IoT forensics to aid cybercrime investigation. These include:

- **Data Locations.** In IoT systems, the data are saved at various locations in dynamic devices that may be regulated by different administrations. Consequently, the investigators undergo serious problems to identify which regulations are to be followed when the device was used to commit a crime [245]. In this type of situation, crime investigation becomes a more complicated task. So, there is a need for standard processes and mechanisms to address this issue.
- **Forensic Automation.** There are numerous technical issues faced during the automated IoT forensic analysis. The major problems which affect the process are the dynamic nature of the devices and the involvement of advanced methods in the process of forensic investigation. To obtain a real-time solution to the problem, there is a requirement for improved IoT automation. The authors in [164], presented a novel direction to IoT forensics by introducing an automated technique for forensics examination. It is also impressed upon by the authors that the diversity of IoT devices is the main hindrance in the real-time implementation of the proposed technique. Therefore, some standard mechanisms are required to deal with the heterogeneity of the devices and collected data.
- **IoT Device Management:** In an IoT environment, sometimes a particular device malfunctions and starts generating malignant data. Although, it may require shutting that device down but may not be feasible for the forensic investigator to do so because of the owner's rationality. For example, if in a smart home a washing machine is initiating vengeful data packets, but the owner does not pass his consent to stop it as it may disturb his daily functioning. This may lead to a big challenge for crime investigator expert. Therefore, a due attention needs to be given to design the required mechanisms to provide the crime investigators freedom for forensic investigation without ceasing the working of things.

- **Forensic Analysis of Data in IoT.** Forensic investigators deal with a large volume of IoT data using various analysis techniques during the process of crime investigation [246]. In an IoT environment, the data are collected and analyzed from various devices and the results are used for various types of decision making [247]. As the process of data analysis and interpretation is complex, the accuracy of the results and further investigation is affected [186]. Therefore, the need for more standardized, simple and accurate data analysis tools and techniques arises.

- **Scope and Life of Digital Forensic Evidence.** The limited storage of IoT devices deters the availability of evidence for a long time which results in the loss of crucial data related to cybercrime[160]. To overcome this problem, forensic data should be transferred frequently to the cloud. However, the process of data transfer gives rise to another challenge to ensure that evidence has not been manipulated during the process. Another major issue is related to the visibility of the evidence. The presence of a few malignant sensors at the crime scene may affect the working of the forensic investigators to locate the witness equipment. Although, log files from various devices may assist the forensic experts but do not provide the complete set of evidence for the investigation.

- **Privacy of the User.** The entanglement of IoT devices in various domains has made human life very comfortable. However, it has put the privacy of the users of smart devices at stake. It has been observed that there is a lack of privacy-specific forensic mechanisms for the IoT environment [248]. The main loophole of most of the available forensic solutions is that the privacy aspect of the users is ignored during the process of investigation[249]. All investigation solutions proposed in [187], [190], [250] have serious privacy challenges. In very diverse and dynamic IoT systems, the practice of suitable privacy measures can enhance the involvement of digital evidence for cybercrime investigations.

- **Security in IoT devices.** The diverse nature of devices in the IoT environment opens a new space for unauthorized users to attack the system which is very difficult to identify during the forensics investigation. Consequently, the process of collection of evidence becomes more tedious. Therefore, it is essential that during the design of various forensic investigation mechanisms, the diverse nature of IoT systems should be kept in mind [251]. The authors introduced the concept of security and privacy in [68], [252]. Proposed approaches and algorithms provide more liberty to forensic investigators by leaving aside security issues. By considering the diverse and dynamic nature of the IoT environment, more such kinds of techniques are the need of the hour in cybercrime investigation [253, 254].

- **Other issues and future research**

  During the study of various challenges, it has been observed that there is a requirement of more standardized techniques and mechanisms to administer the data gathered from heterogeneous and dynamic devices to facilitate the process of cybercrime investigation. Due to the diversity of the formats of the data gathered from the various devices, there is also a requirement of more sophisticated data analysis tools and techniques. Advanced methods need to be proposed to facilitate the investigators with the liberty to work without interrupting the working of smart devices and equipment. As the storage capacity of most of the smart devices is limited; there is a requirement of accurate and efficient techniques to transfer the forensic data from IoT devices to the cloud without any loss of evidence. Suitable measures also need to be practiced ensuring the privacy of the user's personal data during the process of investigation.

## 8. Conclusions

IoT is a developing technology, which has bestowed human life with comfort. However, the growing practice of IoT devices in various domains related to business and personal life has put personal and data security at greater risk. A large volume of data is exchanged openly among the various smart devices in an IoT environment which attracts hackers to penetrate the security system. The dependence of IoT systems on wireless communication technologies makes them prone to cyber-attacks which is the root cause of cybercrime. In this paper, we present the various elements of the IoT framework like architecture, protocols, technologies, and application domains. A detailed review of the security aspects of an IoT environment from the years 2010-2020 is presented. Various security aspects which may facilitate intruders to commit cybercrime are also discussed. Implementation of the security mechanisms at each of the layers of IoT architecture is presented in this survey. The role of IoT Forensics and advanced technologies in cybercrime investigation is impressed upon in this review. This survey also consists of patents reported and real-time applications developed to mitigate the problems occurring due to cybercrime in IoT devices. At last, the various open research challenges to be addressed are discussed to facilitate the process of cybercrime investigation in the IoT systems.

## Data Availability Statement

Any data or material used in the survey is referred to in the article.